\documentclass[twocolumn,times,tighten]{aastex631}

\usepackage[caption=false]{subfig}
\PassOptionsToPackage{mathlines}{lineno}
\usepackage{amsmath}
\usepackage{graphicx}
\usepackage{units}
\usepackage{comment}
\usepackage{verbatim}
\usepackage{hyperref}
\usepackage{array}
\usepackage{xcolor}
\usepackage{lineno}

\newcommand{\Lya}{\hbox{{\rm Ly}\kern 0.1em{$\alpha$}}}
\newcommand{\HI}{\hbox{{\rm H}\kern 0.1em{\sc i}}}
\newcommand{\HeII}{\hbox{{\rm He}\kern 0.1em{\sc ii}}}
\newcommand{\MgII}{\hbox{{\rm Mg}\kern 0.1em{\sc ii}}}
\newcommand{\FeII}{\hbox{{\rm Fe}\kern 0.1em{\sc ii}}}
\newcommand{\CIV}{\hbox{{\rm C}\kern 0.1em{\sc iv}}}
\newcommand{\SiIV}{\hbox{{\rm Si}\kern 0.1em{\sc iv}}}
\newcommand{\OVI}{\hbox{{\rm O}\kern 0.1em{\sc vi}}}
\newcommand{\MgIIdblt}{{\rm Mg}\kern 0.1em{\sc ii}~$\lambda\lambda 2796, 2803$}
\newcommand{\subHI}{\hbox{\tiny{\rm H}\kern 0.05em{\sc i}}}
\newcommand{\subMgII}{\hbox{\tiny{\rm Mg}\kern 0.05em{\sc ii}}}
\newcommand{\subH}{\hbox{\tiny{\rm H}}}
\newcommand{\subMg}{\hbox{\tiny{\rm Mg}}}
\newcommand{\subCIV}{\hbox{\tiny{\rm C}\kern 0.05em{\sc iv}}}
\newcommand{\subU}{_{\hbox{\tiny U}}}
\newcommand{\subL}{_{\hbox{\tiny L}}}

\newcommand{\kms}{km~s$^{-1}$}

\journalinfo{Version dated March 17, 2024}

\shorttitle{\sc The OzDES {\MgII} Survey}
\shortauthors{\sc Abbas, Churchill, Kacprzak, et~al.}

\defcitealias{Churchill_2020}{C20}
\defcitealias{mathes2017}{M17}
\defcitealias{Codoreanu2017}{C17}
\defcitealias{bosman2017}{B17}
\defcitealias{sebastian23}{S23}

\begin{document}

\title{The Mass Density of {\MgII} Absorbers from the Australian Dark Energy Survey}

\author[0009-0006-4626-832X]{Asif Abbas}
\affiliation{Dept of Astronomy, New Mexico State University, Las Cruces, NM 88003, USA}

\author[0000-0002-9125-8159]{Christopher W. Churchill}
\affiliation{Dept of Astronomy, New Mexico State University, Las Cruces, NM 88003, USA}

\author[0000-0003-1362-9302]{Glenn G. Kacprzak}
\affiliation{Centre for Astrophysics and Computing, Swinburne University, Victoria 3122, Australia}
\affiliation{ARC Centre of Excellence for All Sky Astrophysics in 3  Dimensions (ASTRO 3D), Australia}
\author[0000-0003-1731-0497]{Christopher Lidman}
\affiliation{Centre for Gravitational Astrophysics, College of Science, The Australian National University, ACT 2601, Australia
}
\affiliation{The Research School of Astronomy and Astrophysics, Australian National University, ACT 2601, Australia}

\author[0000-0002-9289-7956]{Susanna Guatelli}
\affiliation{University of Wollongong, Centre for Medical Radiation Physics, New South Wales, Australia}
\affiliation{Illawarra Health and Medical Research Institute, University of Wollongong, New South Wales, Australia}

\author[0000-0003-4169-9738]{Sabine Bellstedt}
\affiliation{ICRAR, University of Western Australia, 35 Stirling Highway, Crawley, WA 6009, Australia}

\begin{abstract}
We present an all-southern sky survey for {\MgIIdblt} doublet absorbers in 951 $z < 4$ AGN/quasar spectra from the Australian Dark Energy Survey (OzDES). The spectral resolution ranges from $R = 1400$--1700 over the wavelengths 3700--8800~{\AA}. The survey has a $5\sigma$ detection completeness of 50\% and above for rest-frame equivalent widths $W_r(2796) \geq 0.3$~{\AA}. We studied 656 {\MgII} absorption systems over the redshift range $0.33 \leq z \leq 2.19$ with equivalent widths $0.3 \leq W_r(2796) \leq 3.45$~{\AA}. The equivalent width distribution is well fit by an exponential function with $W_* = 0.76\pm 0.04$~{\AA} and the redshift path density exhibits very little evolution. Overall, our findings are consistent with the large, predominantly northern-sky, surveys of {\MgII} absorbers. We developed and implemented a Monte Carlo model informed by a high-resolution {\MgII} survey for determining the {\MgII} mass density, $\Omega\subMgII$.  We found $\Omega\subMgII \sim 5\times 10^{-7}$ with no evidence of evolution over a $\sim\! 7$~Gyr time span following Cosmic Noon.  Incorporating measurements covering $2.0 \leq z \leq 6.4$ from the literature, we extended our insights into {\MgII} mass density evolution from the end of reionization well past the Cosmic Noon epoch. The presented Monte Carlo model has potential for advancing our knowledge of the evolution of mass densities of metal-ions common to quasar absorption line studies, as it exploits the efficiency of large low-resolution surveys while requiring only small samples from expensive high-resolution surveys.
 
\end{abstract}


\section{Introduction} 
\label{sec:intro}

Characterizing the evolution of the gaseous properties of the universe is a key step to understanding the processes that shape the universe of galaxies and groups and clusters of galaxies \citep[][]{tumlinson_2017_the}. These processes are also responsible for the build up of the chemical elements through what is often called the cosmic baryonic cycle \citep[e.g.,][]{peroux2020}. The study of cosmic gas between, surrounding, and embedded in galaxies is facilitated by the tool of quasar absorption-line spectroscopy. 

Absorption from specific ionization states of different chemical elements probe and provide insight into various gaseous structures and environments in the universe.  For example, absorption from five-times ionized oxygen ({\OVI}) probes the hotter more diffuse regions of the universe \citep[e.g.,][]{cen99, tripp2006, oppenheimer09, nelson18, bradley22}, whereas absorption from singly ionized magnesium ({\MgII}) probes the cooler denser regions nearer to galaxies \citep[e.g.,][]{bb91, sdp94, kacprzak2010, nielsen_2013_magiicat}. Magnesium is an abundant $\alpha$-process element resulting from Type-II supernovae, so it is a good tracer of gaseous environments that has been enriched by recent star formation \citep[e.g.,][]{jtl1996}.

In first decades following the discovery of the cosmological nature of quasars \citep{schmidt63}, {\MgIIdblt} resonant fine-structure absorption lines were found to be a common feature in their spectra \citep[e.g.,][]{kinman67, carswell75, burbidge77}.  Over the next decades, the statistical characteristics of this absorber population were slowly uncovered \citep[e.g.,][]{bergeron84, Lanzetta_87, tytler87, ssb1988, caulet89, steidel1992}. In these early optical studies, the redshifted {\MgII} doublet was observable over the redshift range $0.3 \leq z \leq 2.2$. These works established that {\MgII} absorbers were cosmologically distributed and examined their redshift evolution, including their redshift path densities, redshift clustering, and equivalent width and doublet ratio distributions. They were mostly sensitive to absorbers with rest-frame equivalent widths $W_r \geq 0.3$~{\AA}. The higher-resolution survey of \citet{petitjean1990} was able to examine the kinematics and multi-component structure, and their velocity splitting and column density distributions. 

With the advent of Sloan Digital Sky Survey (SDSS), successively larger optical low-resolution ground-based surveys were undertaken \citep[e.g.][]{nestor_2005_mgiiabsorption, lundgren2009,  Seyffert_13, zhu_2013_the, Raghunathan2016}, from which the statistics of the absorber characteristics were measure to high precision. The advent of optical high-resolution spectrographs on 10-meter class telescopes opened two new windows: (1) the sensitivity was improved such that the characteristics of absorbers with $W_r \leq 0.3$~{\AA} could also be studied \citep[e.g.,][]{churchill_1999_the, narayanan07, mathes2017}, and (2) the kinematics and component structure of the absorbers could be studied 
\citep[e.g.,][]{cv01, churchill03, prochter2006, mshar07, Churchill_2020}.  And finally, the advent of sensitive infrared spectrographs on large telescopes allowed {\MgII} absorber studies to be extended up to $z \sim 7$ \citep[e.g.][]{matejek-simcoe2013, chen-simcoe2017, Codoreanu2017, Davies2023-Absorbers, sebastian23}. These moderately high-resolution spectra allowed analysis provided both the kinematics and the statistical characteristics of absorbers. 

Though the characterization of the nature of {\MgII} absorbers has continually improved \citep[see][for a brief summary]{churchill2024}, the cosmic evolution of the mass density of {\MgII} absorbing ions, $\Omega\subMgII$, is not well constrained across all redshifts over which {\MgII} has been studied.
Recently, \citet{sebastian23} obtained robust measurements of $\Omega\subMgII$ over the redshift range $2 \leq z \leq 6.5$ using the high-resolution X-shooter spectrograph \citep{vernet2011-xshooter} on the Very Large Telescope.  These measurements represent a significant improvement over the similar work by \citet{Codoreanu2017}. However, measurements of $\Omega\subMgII$ for the redshift range $0.3 \leq z \leq 2$ are only lower limits \citep{mathes2017}. The reason they are lower limits is because the column densities were approximated using the apparent optical depth method, which is known to provide lower limits when absorption profiles exhibit unresolved saturation \citep{savage_aod91, jenkins1996}.

Measuring $\Omega\subMgII$ is challenging because it requires an accurate characterization of the {\MgII} column density distribution, and that requires Voigt profile fitting of {\MgII} absorbers in high-resolution spectroscopy. First, high resolution spectroscopy is expensive, requiring time on competitive 10-meter class telescopes. Second, there are only so many bright quasars suitable for high signal-to-noise high-resolution spectroscopy. Third, Voigt profile fitting complex absorption systems is time intensive work. These bottlenecks limit the sample sizes for accurately measuring $\Omega\subMgII$. To date, the SDSS has archived roughly $10^6$ low-resolution quasar spectra containing roughly $10^5$ {\MgII} absorbers \citep[e.g.,][]{Raghunathan2016, Anand2021}. We ask, is it plausible to develop a method to accurately estimate the column densities of these absorbers and exploit the existing huge archives to obtain high precision estimates $\Omega\subMgII$?

As previously mentioned, for $2 \leq z \leq 7$ recent works \citep{Codoreanu2017, sebastian23} have measured $\Omega\subMgII$ using the X-shooter.  \citet{sebastian23} found that $\Omega\subMgII$ increases by a factor of 10 to 100 from $z \sim 7$ to $z \sim 2$.  This suggests an increase in $\alpha$-group elements in low-ionization gas structures in the epoch leading up to Cosmic Noon \citep[e.g.,][]{forster20}, when the universe is at its peak activity ($2\leq z \leq 3$).  On account that  \citet{mathes2017} measured lower limits on $\Omega\subMgII$ for $z<2$, the evolution of the mass density of {\MgII} absorbers following Cosmic Noon remains unknown.

In this paper, our main goals are to present a census of the {\MgII} absorbers in the OzDES quasar spectra database and to develop and pilot a method to better constrain $\Omega\subMgII$ for $z<2$ using low-resolution quasar spectra. In addition to providing deeper insight into the evolution of {\MgII} absorbers, we may be able to ultimately leverage the existing large archives of low-resolution spectra to obtain precision measurements of mass densities of commonly absorber populations, such as {\MgII}, {\FeII}, {\CIV}, {\SiIV}, etc. Our secondary goal is to conduct a blind and unbiased survey of {\MgII} absorbers exclusive to the southern hemisphere.  For both goals, we use the archived low-resolution AGN/quasar spectra obtained through the Australian Dark Energy Survey \citep[OzDES,][]{yuan2015, childress_17, lidman_2020_ozdes}.  

In Section~\ref{sec:algo}, we present the AGN/quasar catalog and the data reduction.
In Section~\ref{sec:thesurvey}, we detail the absorption lines survey methods, the detection algorithm, and assess the completeness of the algorithm. Section~\ref{sec:Results} presents the survey results, including the equivalent width distribution, and the path density of the absorption systems. We also provide a brief comparison to northern hemisphere studies. In Section~\ref{sec:massdensity}, we introduce a Monte Carlo model to estimate the mass density of {\MgII} absorbers. In section~\ref{sec:discussion}, we compare higher redshift measurements and examine possible evolution in the mass density across the Cosmic Noon epoch.  We conclude in Section~\ref{sec:conclusion}. Throughout this paper, we adopt $\Omega_M = 0.321$, $\Omega_\Lambda = 0.679$ and $H_0=71.9$~{\kms}~Mpc$^{-1}$ from the \citet{Planck2018}.

\section{Data and Data Reduction} 
\label{sec:algo}

The Dark Energy Survey \citep[DES,][]{flaugher2005} is a ground-based optical and near infrared survey designed to improve our understanding of the accelerating expansion of the universe and the nature of dark energy. The Australian Dark Energy Survey \citep[OzDES,][]{yuan2015, childress_17, lidman_2020_ozdes} is a ground-based optical survey and is the primary source of galaxy/AGN spectroscopy for DES. 

\subsection{Australian Dark Energy Survey}

The OzDES survey used the 2-degree field (2dF) fiber positioner and the AAOmega spectrograph \citep{OzDES_AAOmega} on the 3.9-meter Anglo-Australian Telescope (AAT). OzDES targeted 10 deep fields in the DES supernova survey over 6 years to obtain the redshifts of thousands of galaxies. OzDES ran 6 observing seasons from 2013 to 2019, with two extra deep fields, C3 and X3, receiving the highest priority, followed by two E fields, E1 and E2. These fields were observed repeatedly enabling redshifts to be obtained for faint sources. Of all the sources observed, AGN received a quarter of the total fiber hours allotted to OzDES. 

The AGN/quasars selected for observation were in the redshift interval $0 \!<\! z \!<\! 4$. The 951 AGN/quasars were monitored with the goal of measuring the lags between the continuum from the accretion disk and the broad lines from the broad line emission regions \citep[e.g.,][]{yu21, yu23, penton22, malik23}. The AGN were observed with the x5700 dichroic, resulting in a spectral range from 3700~{\AA} to 8800~{\AA} at a resolution ranging from $R=1400$--1700.

\subsection{Data Reduction/Calibration}

All OzDES quasar spectra were processed through an upgraded version of the analysis package {\tt 2dfdr} \citep{croom2004}. As described in more detail in 
\citet{yuan2015} and \citet{childress_17}, these steps include (1) overscan and bias subtraction, bad pixel masking, and preliminary cosmic ray removal, (2) tram line mapping, i.e., locating fiber traces, (3) spectral extraction, (4) wavelength calibration in units of constant wavelength, (5) sky subtraction using extracted sky fibers and a principle component analysis to remove sky line residuals, and (6) combining and splicing, which involves co-addition of multiple frames and cosmic ray removal. The OzDES spectra are resampled to a common wavelength grid. This is done when the spectra from the red and blue arms of the spectrograph are combined. The redshifts of the quasars were determined by interactive software and visual inspected by OzDES team members using the interactive redshifting software RUNZ.

\begin{figure}[ht] 
\centering
\includegraphics[width=0.98\columnwidth]{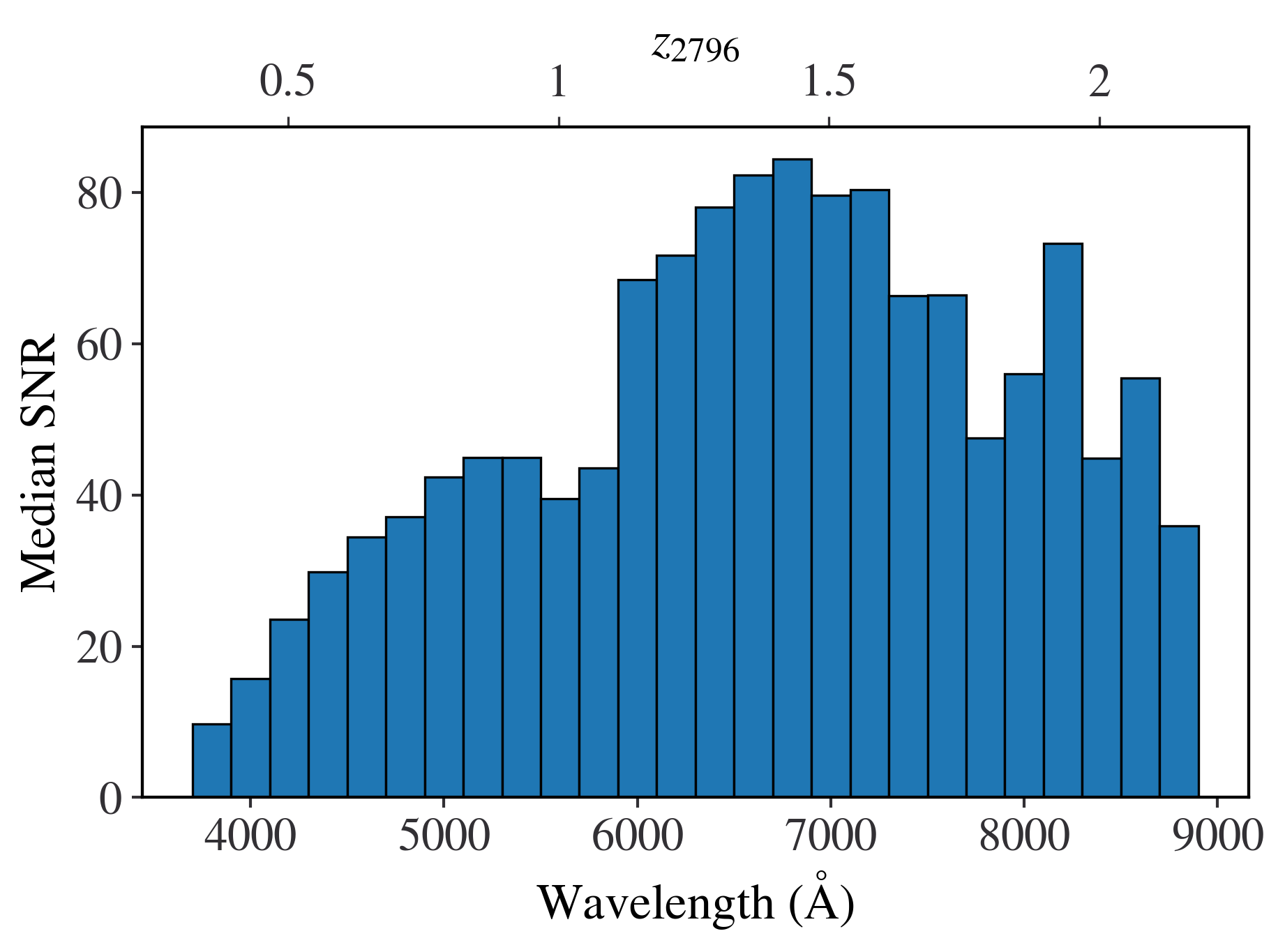}
\captionsetup{justification=justified, singlelinecheck=off} 
\caption{A histogram of the median signal-to-noise ratio (SNR) of the 951 OzDES quasar spectra in 200~{\AA} bins. (lower axis) SNR as a function of wavelength covering the spectral range of the spectra, 3700 to 8900~{\AA}. (top axis) SNR as a function of the redshift for the {\MgII}~$\lambda 2796$ line. The dip at 5700~{\AA} ($z_{2796} \sim 1$) is due to the x5700 dichroic.
}
\label{fig:SNR}
\end{figure}

In Figure~\ref{fig:SNR}, we show the median signal-to-noise ratio (SNR), in 200~{\AA} bins, from 3700~{\AA} to 8900~{\AA} for all 951 OzDES quasar spectra. The median SNR is less than 25 below 4200~{\AA}, making it difficult to reliably identify {\MgII} absorption lines below $z \simeq 0.43$. Therefore, we limit our search for absorption lines to above 4200~{\AA}. We further restrict our search for {\MgIIdblt} doublets redward of the {\Lya} line in order to avoid confusion in the {\Lya} forest.

\begin{figure*}[hbt] 
\centering
\includegraphics[width=\textwidth]{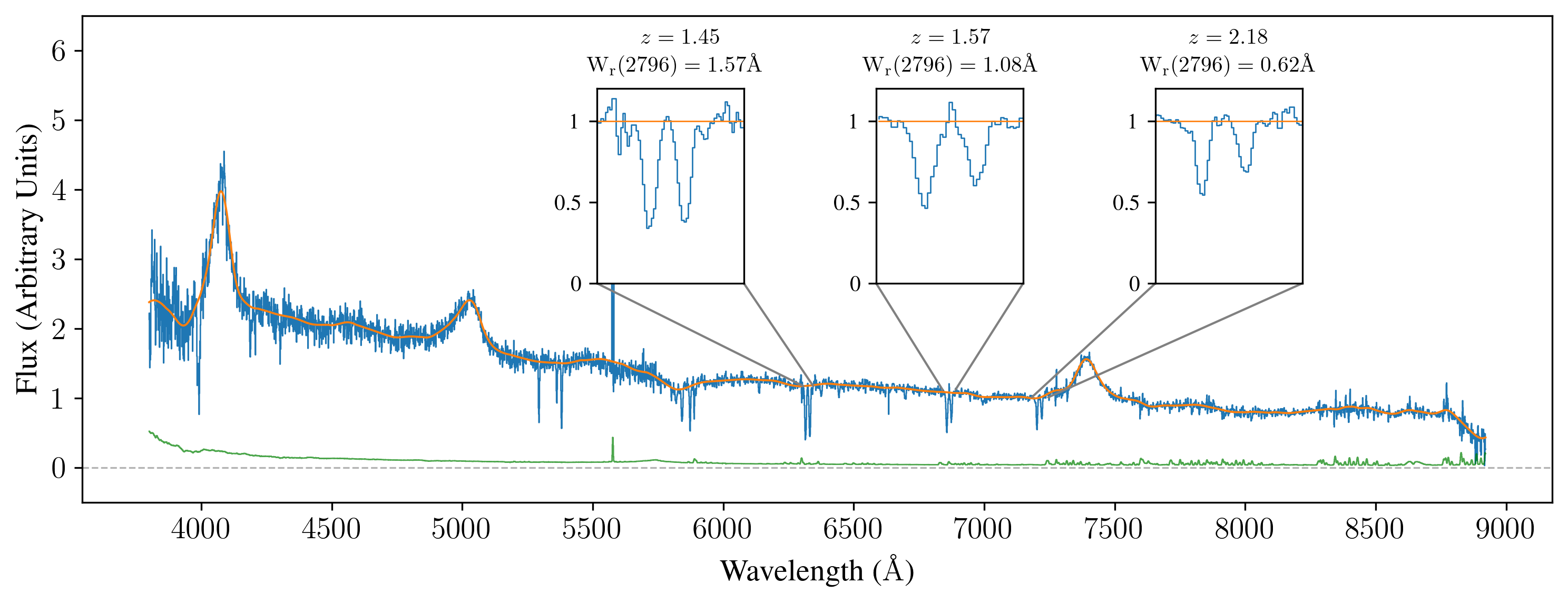}
\captionsetup{justification=justified, singlelinecheck=off} 
\caption{OzDES spectrum of the quasar SVA1-COADD-2925403880 ($z\sim 2.41$) showing the continuum model (orange) and the sigma spectrum (green).  Three intervening {\MgII} absorbers are highlighted (insets) at absorption redshifts $z = 1.45$, ($\sim\! 6300$~{\AA}), $z = 1.57$, ($\sim\! 6850$~{\AA}), and $z = 218$, ($\sim 7200$~{\AA}).}
\label{fig:example spectrum}
\end{figure*}

In Figure~\ref{fig:example spectrum}, we show a typical quasar spectrum from the OzDES survey.  This quasar resides at $z\sim 2.41$. There are three intervening {\MgII} absorbers detected in the spectrum.  The continuum fit is given by the orange curve.

\subsection{Continuum Fitting} 
\label{sec: Continuum}

To estimate the continuum for each spectrum, we employ a three-step process. In Step~1, we approximate the continuum at a pixel $j$ as the median flux in a window centered around the pixel $j$. However, the median continuum is not accurate across absorption and emission lines. For Step~2, we identify all statistically significant (greater than $3\sigma$) absorption features using the methods described in \citet{Lanzetta_87}. In Step~3,  we replace the continuum across absorption and emission lines with Legendre polynomial fits using the method outlined in \citet{sembach_savage_1992}. In this step, the absorption features are masked from the fits.  This process rectifies the systematic errors present in the median continuum and provides a  formal framework for accounting for of uncertainties due to the estimated continuum flux across absorption lines (see \ref{sec:Leg Poly Fit}).  In the following subsections we detail the continuum fitting.

\subsubsection{Step 1: Median Continuum}
\label{sec:step1}

For computing the median flux, we adopt a 50-pixel wide window centered on a pixel $j$. The median centered on pixel $j$ is repeated for pixels $j+1$, $j+2$ and so forth.  At the blue end and red ends of the spectrum, we adopt the median computed for the last pixels for which the full 50-pixel window can be employed. As illustrated in Figure~\ref{fig:continuum}(a), the median tends to be underestimate the continuum across absorption features. We thus need to refine the continuum estimate across the absorption features.

\subsubsection{Step 2: Feature Masking for Refinement} 
\label{sec: Aperture Method}

In order to refine the continuum estimate across absorption features, we first build a catalog of the absorption features. Then, as described in Section~\ref{sec:Leg Poly Fit}, we mask the pixels comprising the features and refine the continuum fit. At this stage in our analysis, these absorption features are not used for identifying {\MgII} doublets; they are used only for pixel masking and continuum fit refinement.

To locate the features, we use a slightly modified version of the ``aperture method'' as formulated by \citet{Lanzetta_87}. This method compares the equivalent width computed over a fixed number of pixels (the aperture) to its uncertainty and identifies spectral regions where the equivalent width in an aperture is significant.
 
We define the equivalent width of pixel $j$ as $w_j = D_j\Delta\lambda_j$, where $D_j = 1 - {f_j}/{f_j^c}$ is the flux decrement in the pixel, and where $\Delta\lambda_j = \lambda_{j+1} - \lambda_j$. The quantity $f_j$ is the flux density and $f_j^c$ is the estimated continuum at pixel $j$ from Step~1. The uncertainty in the equivalent width per pixel is $\sigma_{w_j} = \sigma_{D_j}\Delta\lambda_j$. At this stage, we assume no uncertainty in the continuum estimate, so that $\sigma_{D_j} = |\sigma_{f_j}|$.

To identify candidate features, we scan the spectrum from blue to red for the condition $w_j/\sigma_{w_j}\geq3$, which we deem a significant equivalent width for pixel $j$.  We then search for the blueward and redward extremes of the candidate absorption feature by scanning the spectrum blueward and redward of pixel $j$ for the first pixel satisfying the condition $w_j/\sigma_{w_j}\geq 1$. This yields pixel $j^-$ for the blue extreme of the feature and pixel $j^+$ for the red extreme of the feature. The region from  pixel $j^-$ to  pixel $j^+$ comprises the aperture over which the candidate feature is masked from the refinement to the continuum fit.
 
\subsubsection{Step 3: Legendre Polynomial Fit} 
\label{sec:Leg Poly Fit}

The existing median continuum spanning an absorption feature is replaced with a Legendre polynomial fit. To ensure a robust fit, the first 150 ``clean pixels" (pixels devoid of statistically significant absorption) on both sides of a feature are included in the fit. This step yields a refined continuum across absorption features.
 
Following \citet[][]{sembach_savage_1992}, we adopt Legendre polynomials due to their orthonormal properties. Being orthogonal, the fitted coefficients for each order of the polynomial are independent of one another, which has the added property that the uncertainties in the coefficients are also independent (the covariance matrix is diagonal).  As we show below, this allows us to objectively determine the most optimal order, $m$, of the fitted polynomial.  

A Legendre polynomial of order $m$ is written
\begin{equation}
    P_m(x) = \sum_{k=0}^m a_kP_k(x)\,.
\label{eq:LegendrePoly}
\end{equation}
where $P_0(x) = 1$, $P_1(x) = x$, and 
\begin{equation}
    P_k(x) = [(2k\!-\!1)xP_{k\!-\!1}(x) - (k\!-\!1)P_{k\!-\!2}(x)]/k \, ,
\end{equation}
for $k\geq2$, where $k$ is the index of each term of the polynomial. To enable the orthonormal properties, for each pixel $j$ the $P_k(x_j)$ are defined over the range $x_j \in [-1,+1]$, where we adopt the definition $x_j = (\lambda_j - \bar{\lambda})/ \delta \lambda$, where $\lambda_j$ is the wavelength at pixel $j$, and where 
$\bar{\lambda} = (\lambda_+ + \lambda_-)/2$ and $\delta \lambda = (\lambda_+ - \lambda_-)/2$ for the 300 pixel interval over which $\lambda_- \leq \lambda_j \leq \lambda_+$, with $\lambda_-$ and $\lambda_+$ being the starting and ending wavelengths of the pixel interval.
 
We adopt modified $\chi^2$ least-squares minimization technique, where 
\begin{equation} 
\chi^2_m = \sum_{j=1}^{n_{\rm pix}} \frac{\left(f_j(x_j) - P_m(x_j)\right)^2}{\sigma^2_{f_j}({x_j})}\,,
\label{eq:chi2-cont}
\end{equation}
for order $m$ where $f_j(x_j)$ is the flux at pixel $j$, ${\cal P}_m(x_j)$ is the value of the fitted Legendre polynomial of order $m$ evaluated at pixel $j$, and $\sigma^{}_{f_j}({x_j})$ is the uncertainty in the flux at pixel $x_j$, and $n_{\rm pix}$ is the number of pixels in the fitted interval of the spectrum. For the least-squares fitting, we use the LEGFIT\footnote{numpy.polynomial.legendre.legfit (NumPy v1.26 Manual).} function from the NumPy module. This function returns both the coefficients and the sum of squared residuals of the least squares fit.

The lower the order, $m$, the fewer the fitting coefficients and the smaller the formal error in the continuum fit, $P_m(x_j)$. Thus, it is desirable to use the smallest order polynomial that provides a statistically significant improvement in the refinement of the continuum. The orthonormal properties of Legendre polynomials provides the desirable property that adding or removing the highest-order term from the fit does not impact the fitted values of the fitting coefficients for the other lower-order terms.  

\begin{figure*}[ht]
\centering
\includegraphics[width=0.85\textwidth]{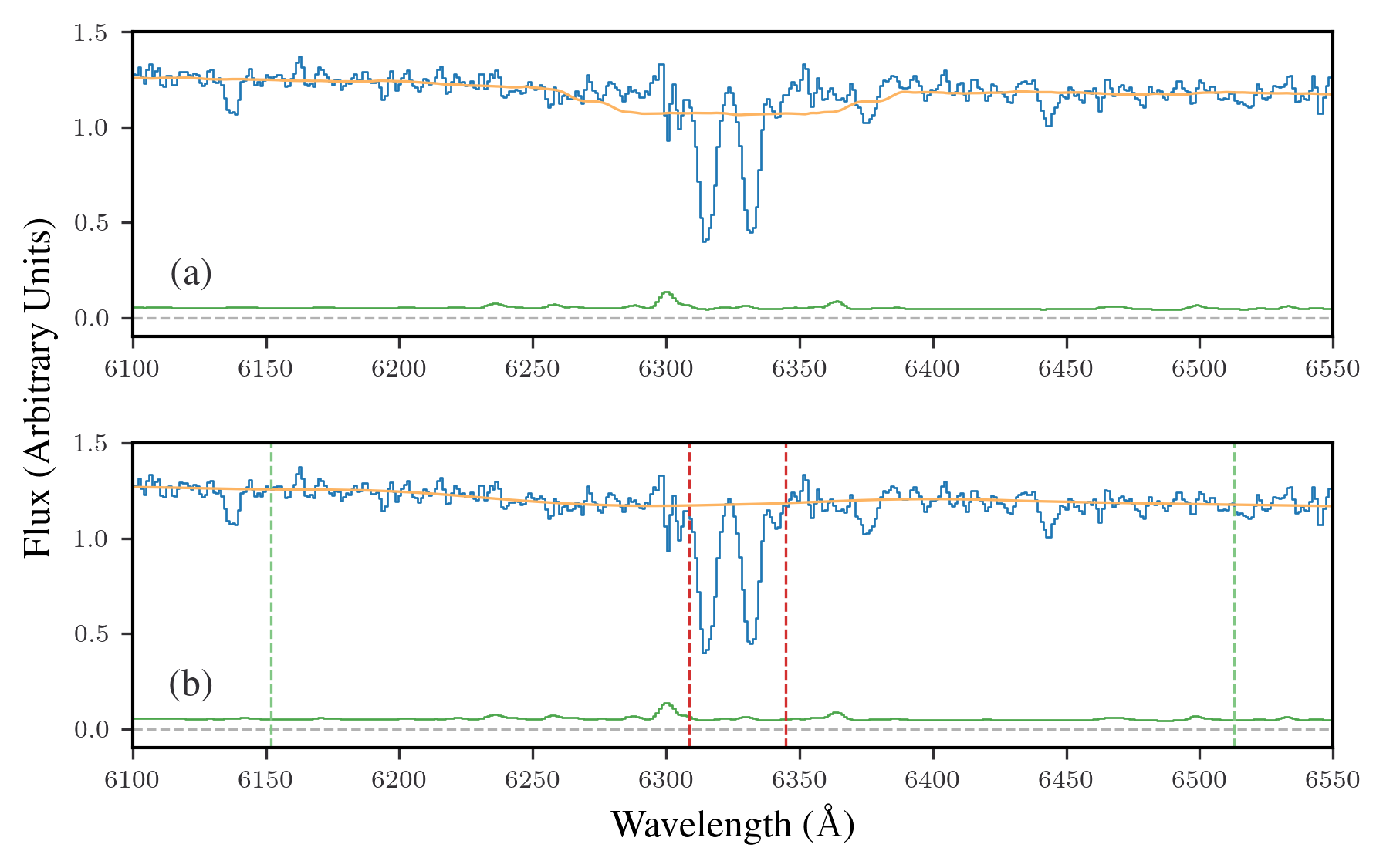}
\captionsetup{justification=justified} 
\caption{
Expanded region highlighting the $z=1.45$ {\MgII} absorber in the SVA1-COADD-2925403880 spectrum. (a) The continuum estimated  (orange curve) using the median continuum method described in Section~\ref{sec:step1} The sigma spectrum is shown in green.  The absorption feature causes an underestimation in the local continuum. (b) The feature is objectively identified and its aperture (vertical red lines) is determined using the aperture method described in Section~\ref{sec: Aperture Method}. A region spanning $\pm 150$ pixels is defined (vertical green lines) across which a Legendre polynomial fit is applied as described in Section~\ref{sec:Leg Poly Fit}.}
\label{fig:continuum}
\end{figure*}

We use the $F$-test to determine the optimal order of the fit. By optimal, we mean the minimum $m$ required to provide a statistically robust fit in that adding an additional order does not statistically improve the $\chi^2$ statistic.  We aim to obtain the most statistically significant fit while minimising the formal error in the fit. The $F$-test statistic for comparing the fit of order $m$ to order $m-1$ is 
\begin{equation} 
F_m = \frac{\chi^2_{m-1} - \chi^2_m}{\chi^2_m/\nu_m} \, ,
\label{eq:F-cont}
\end{equation}
where $\nu_{m} = n_{\rm pix} - m - 1$ is the number of degrees of freedom for a fit of order $m$ applied to $n_{\rm pix}$. 

An additional term is retained only if the probability ($p$-value) that obtaining $F_m$ greater than the measured value is smaller than a specified value, which is $p=1-{\rm CL}$, where CL is the double-sided confidence level. 
The $p$-value is then computed using the regularized incomplete beta function,
\begin{equation}
  p = 2 \frac{\Gamma(a+b)}{\Gamma(a)\Gamma(b)}\int_{0}^{x} t^{a-1} (1-t)^{b-1} dt \, ,
\label{eq:p-cont}
\end{equation}
where $a = \nu_{m}/2$, $b = \nu_{m-1}/2$, and where $x = \nu_m / (\nu_m \!+\! \nu_{m-1}F_m)$.  We adopt a confidence level of 95\%, so that an additional term is only retained in the fit when $p < 0.05$.  That is, when a term corresponding to order $m+1$ is added and $p \geq 0.05$, that term is rejected and we adopt order $m$. Almost all of the polynomial fits are of the order 4--6.

Once the order is determined, we can compute the uncertainty in the continuum fitted value at pixel $j$ from
\begin{equation}
\sigma^2_{f^c_j}(x_j) = \sum_{i=0}^{m}\sum_{k=0}^{m}\sigma^2_{ a_{ik}} P_j(x_j) P_k(x_j) \,.
\label{eq:Legendre-sigma}
\end{equation}
where $\sigma^2_{a_{ik}}$ are in the uncertainties in the fitted coefficients obtained from the diagonal of the covariance matrix. Though the corresponding wavelength at pixel $j$ can be computed from
\begin{equation}
\lambda_j = x_j \delta \lambda + \bar{\lambda} = [\lambda_+(1\!+\!x_j) + \lambda_-(1\!-\!x_j)]/2 \, , 
\end{equation}
pixel index $j$ immediately provides the wavelength $\lambda_j$ corresponding to $x_j$. Further details of the applications of Legendre polynomials and the $F$-test can be obtained in Appendix~1.1 of \citet{sembach_savage_1992}.
 
In Figure~\ref{fig:continuum}, we show the continuum estimation across a {\MgIIdblt} doublet with the aperture defined by the red vertical dashed lines. Figure~\ref{fig:continuum}(a) illustrates the dip in the continuum when using the median flux for estimation (Step 1). In Figure~\ref{fig:continuum}(b), following Steps 2 and 3, we show the final adopted continuum normalized spectral region.  The depression has been corrected by applying a 9$^{\rm th}$ order Legendre polynomial fit to 150 ``clean'' pixels on each side of the {\MgII} doublet (those between the green vertical lines, omitting those between the orange and red lines). The optimal order of the polynomial was estimated through the $F$-Test given in Eq.~\ref{eq:F-cont}. 

\section{The Survey}  
\label{sec:thesurvey}

In this section, we describe how we define our detection threshold sensitivity, how we objectively survey for and define candidate {\MgII} doublets, and how we evaluate the completeness of our survey for these absorption lines. 

\subsection{The Detection Threshold}
\label{sec:ISF Weighted}

The aperture method utilized in Section \ref{sec: Aperture Method} is suitable for locating strong absorption features for the purpose of refining the continuum model; however, it lacks sensitivity for detecting weak, unresolved absorption lines. 
Thus, for objectively finding {\MgII} doublets, we adopt the optimized method of \citet{schneider_93}, which employs weighting of the instrumental line spread function (ISF) to maximize the detection sensitivity for absorption. This optimized method is applied following the refinement of the continuum model using the Legendre polynomials.  
 
We modeled the ISF as a Gaussian function with root-mean square width (Gaussian standard deviation)
\begin{equation}
    \sigma_j = \frac{\lambda_j}{2\sqrt{2\ln(2)}R} \, ,
\label{eq:LSFwidth}
\end{equation}
centered on pixel $j$ using the discretized expression with index $n$
\begin{equation}
    {\cal P}_n = \frac{\exp\left(-y_{kj}^2\right)}{\sum_{n=1}^N\exp\left(-y_{kj}^2\right)} \, ,
\end{equation}
where 
\begin{equation}
y_{kj} = \frac{\lambda_j - \lambda_k}{\sigma_j} \, , 
\end{equation}
and where $N = 2J_0 + 1$ is the number of pixels spanning the ISF, where $J_0 = 2p$ with $p=3$ being the number of pixels per resolution element, which yields $N=13$. The convolution index at pixel $j$ is $k = j + (n - 1) - J_0$, which is a function of the ISF index $n$.  The resolution ranges from $R=1400$ in the blue to $R=1700$ in the red. We assumed a linear progression over the wavelength range 3700 to 8900~{\AA}, yielding $R= 1400 + 0.058(\lambda_j-3700)$.

The ISF weighted equivalent width per resolution element centered on pixel $j$ is
\begin{equation}
    w_j = \frac{\Delta\lambda_j}{{\cal P}^2}\sum_{n=1}^N{\cal P}_nD_k \, ,
\quad
   {\cal P}^2 = \sum_{n=1}^N{\cal P}_n^2 \, ,
\label{eq:schneider-w}
\end{equation}
where $D_k$ is the flux decrement defined in Section~\ref{sec: Aperture Method}, and $\Delta \lambda_j$ is the width of pixel $j$. The uncertainty in $w_j$ is the $1\sigma$ limiting equivalent width detection threshold sensitivity per resolution element centered at pixel $j$ in a given quasar spectrum; it is written 
\begin{equation}
    \sigma_{w_j} = \frac{\Delta\lambda_j}{{\cal P}^2}\left(\sum_{n=1}^N{\cal P}_n^2\sigma_{D_k}^2\right)^{1/2} \, .
\label{eq:schneider-wsig}
\end{equation}
The uncertainty in the flux decrement,  
\begin{equation}
\sigma_{D_k}^2 = 
\left[ \frac{f_k}{f^c_k} \right]^2
\left(
\left[ \frac{\sigma^{}_{f_k}}{f_k} \right]^2
+
\left[ \frac{\sigma^{}_{f^c_k}}{f^c_k} \right]^2
\right) \,
\end{equation}
includes the uncertainty in the continuum model.  

In spectral regions where we have refined the continuum estimate around absorption features using the Legendre polynomial fits, we have $f^c_k = P_m(\lambda_k)$ given by Eq.~\ref{eq:LegendrePoly} and $\sigma^{}_{f^c_k}$ given by Eq.~\ref{eq:Legendre-sigma}. As we have adopted the lowest order $m$ that provides a statistically significant refinement in the continuum, we have minimized $\sigma^{}_{f^c_k}$ and thus maximized our detection sensitivity, $\sigma_{w_j}$, while properly accounting for uncertainty in the continuum estimate (see Eqs.~\ref{eq:chi2-cont}, \ref{eq:F-cont}, and \ref{eq:p-cont}). Note that a smaller value of $\sigma_{w_j}$ translates to a smaller (more sensitive) equivalent width detection threshold.  In spectral regions where we did not refine the original continuum estimate, which was based on the median method described in Section~\ref{sec:step1}, we adopt $\sigma^{}_{f^c_k} = 0.4\sigma_{f_k}$, based on the findings of \citet{sembach_savage_1992} and \citet{churchill_2015}.  

\subsection{Locating Candidate Doublets}
\label{sec:locatingdoublets}

Absorption features are objectively identified by scanning each spectrum pixel by pixel searching for  the condition $w_j > N_\sigma \sigma_{w_j}$, where we adopt $N_\sigma = 5$.  In other words, when the equivalent width per resolution element centered on pixel $j$ exceeds or equals $N_\sigma$ times its uncertainty, we consider this to be, at a minimum, a detection of unresolved absorption.  Having identified this pixel, we then search blueward to determine $j_-$ corresponding to the pixel with the first occurrence of $w_j \leq \sigma_{w_j}$ and redward to determine $j_+$, the pixel corresponding to the first occurrence of $w_j \leq \sigma_{w_j}$. This provides the detection aperture of the absorption feature.

\begin{figure*}[ht] 
\centering
\includegraphics[width=0.5\textwidth]{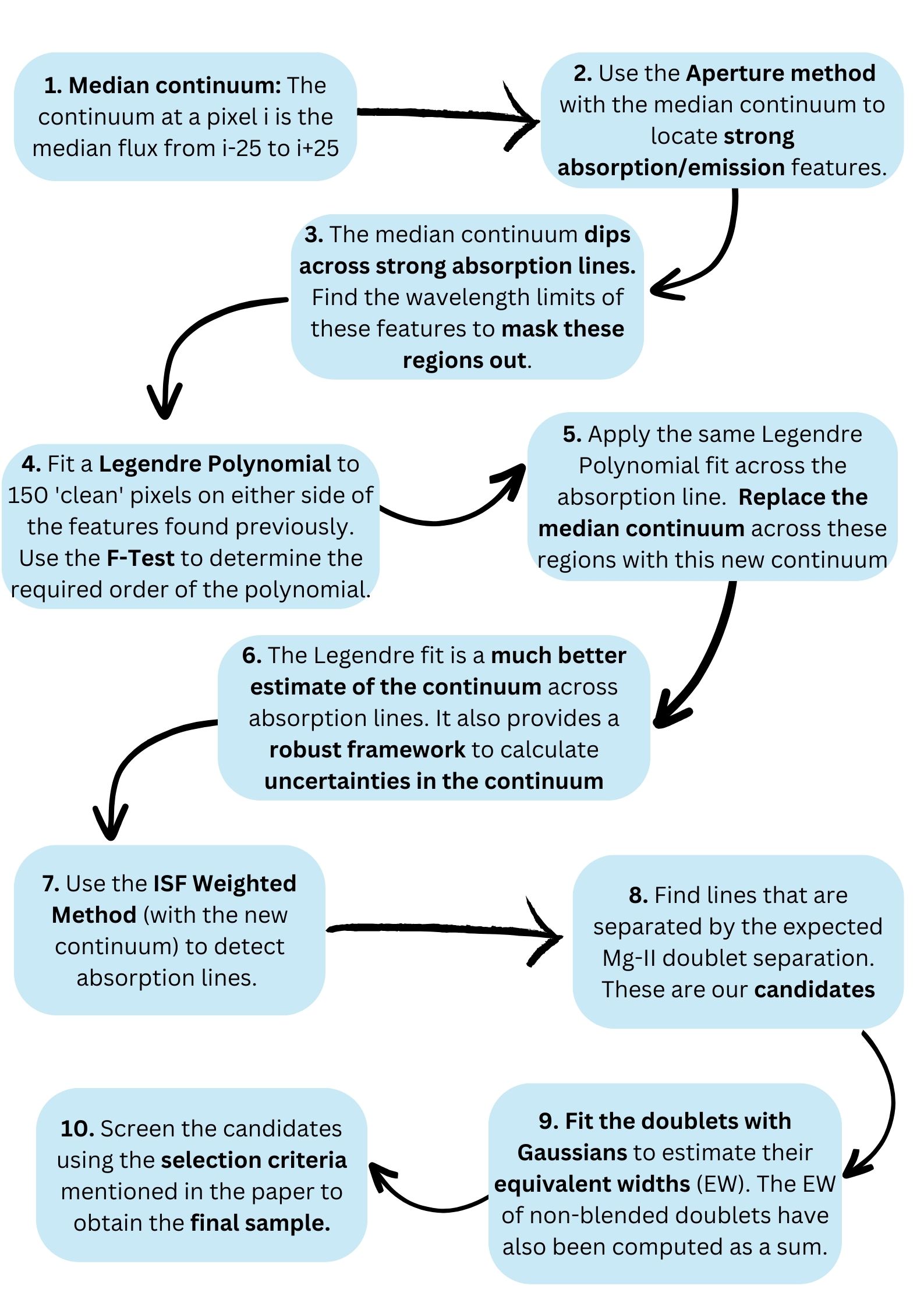}
\includegraphics[width=0.72\columnwidth]{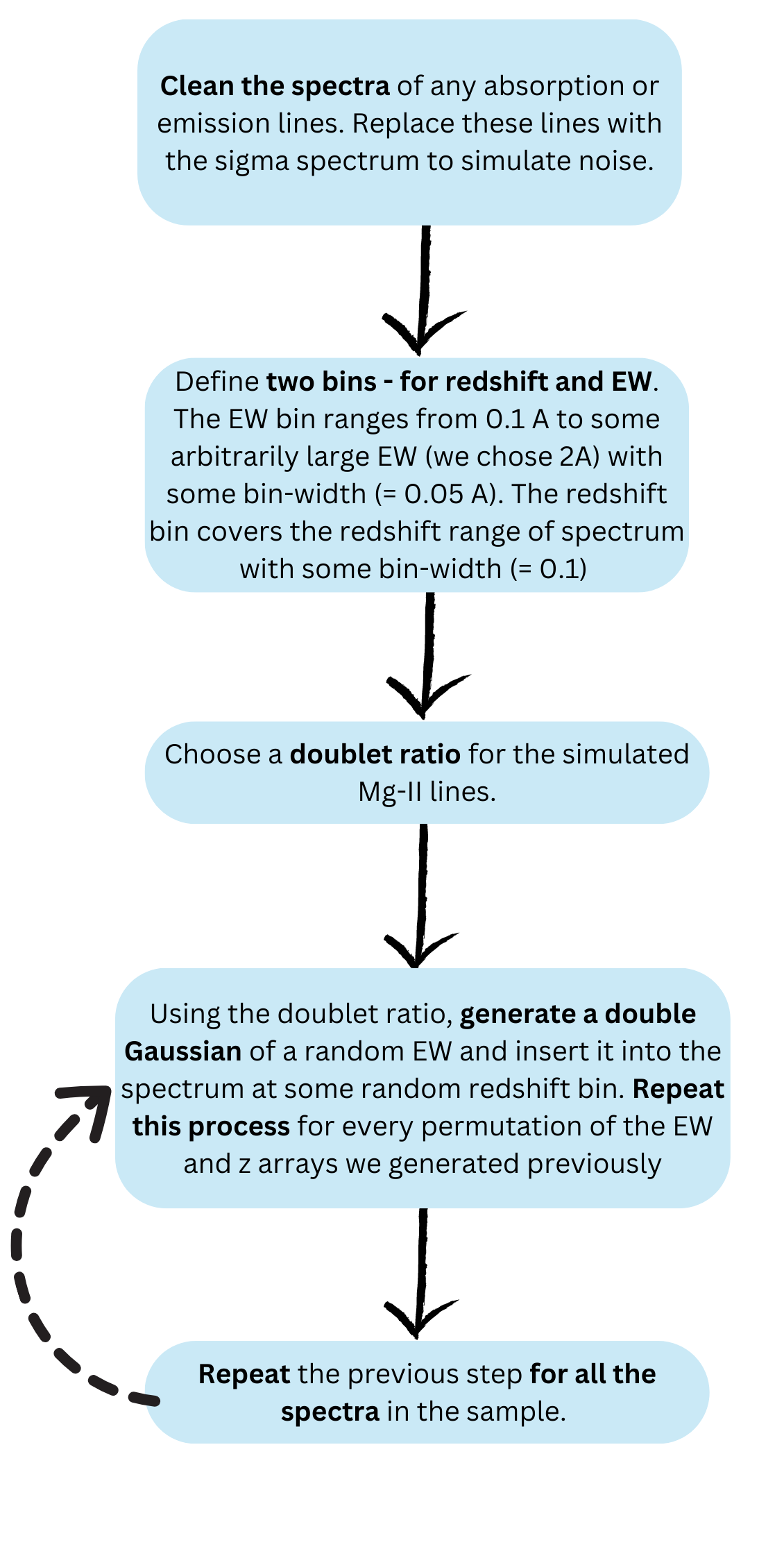}
\captionsetup{justification=justified, singlelinecheck=off} 
\caption{(left) Schematic flowchart summarizing the {\MgII} doublet detection algorithm (Sections \ref{sec: Continuum} through \ref{sec:ISF Weighted}), from modeling the continuum to computing the equivalent width. (right) Schematic flowchart of the steps for computing the completeness function (see Section \ref{sec: Detection completeness}) of the detection algorithm. The values in the parentheses can be altered as required. }
\label{fig:flowchart}
\end{figure*}

We next aim to ascertain if this absorption feature is plausibly one member of a {\MgII} doublet or comprises both members of a partially blended {\MgII} doublet. Within the aperture from $j_-$ to $j_+$, we first locate the number of minima in $w_j$ and their pixel locations. For candidate testing, if there is a single minimum we assume this absorption feature corresponds to a {\MgII}~$\lambda2796$ line centered on the wavelength corresponding to the pixel of the minimum $w_j$.  We then locate the pixel, $j'$, corresponding to the wavelength of the $\lambda 2803$ member of the doublet and determine if the condition $w_{j}/\sigma_{w_{j}}>3$ is met for any of the pixels $j=j'\!-\!1, j', j'\!+\!1$.  If the condition is met, we then determine $j'_-$ and $j'_+$ for the $\lambda 2803$ candidate and check that $j'_- \geq j_+$.  If that latter condition is not met, then we merge the aperture for the two candidate lines into a single aperture covering both lines, meaning we adopt pixel $j_-$ for the blue extremity and adopt pixel $j'_+$ for the red extremity.

To measure the line centers and equivalent widths of the candidate double absorption lines, we simultaneously fit the two lines using a two-component Gaussian function of the form
\begin{equation} \label{eqn: Double Gaussian}
    f(\lambda) = \sum_{i=1}^2 A_i\exp\left\{
    -\frac{(\lambda-\lambda_{c,i})^2}{2\Sigma_i^2}\right\} \, ,
\end{equation}
where we denote $i=1$ for the $\lambda 2796$ transition and $i=2$ for the $\lambda 2803$ transition, and where $A_i$ is the component amplitude, $\lambda_{c,i}$ is the component central wavelength, and $\Sigma_i$ is the line root-mean square width of the component. The fitting was performed  using LMFIT \citep{lmfit}, a non-linear least square minimization curve-fitting package. If a component was fitted with a width smaller than the ISF width, given by Eq.~\ref{eq:LSFwidth}, then the component width was frozen at the ISF width and the minimization was repeated to obtain the component amplitude and central wavelength.  The LMFIT package also yields the uncertainties in the fitted parameters,
$\sigma_{A_i}$, $\sigma_{\lambda_{c,i}}$, and $\sigma_{\Sigma_i}$.

The equivalent width of the each component is computed from the fitted parameters by integrating across the component, which yields,
\begin{equation} 
W_i = \sqrt{2\pi}A_i\Sigma_i 
\label{eq:Gauss-EW}
\end{equation}
with fractional uncertainty
\begin{equation} 
\frac{\sigma_{W_i}}{W_i} = \left[
\left( \frac{\sigma_{A_i}}{A_i} \right) ^2 
+ \left( \frac{\sigma_{\Sigma_i}}{\Sigma_i} \right)^2 
- 2\frac{\sigma_{A_i}\sigma_{\Sigma_i}}{A_i \Sigma_i} 
\right] ^{1/2} \, . 
\end{equation}

To confirm a {\MgII} doublet candidate, we then apply two tests.  First, we examine if the members of the candidate doublet are at a consistent redshift by applying the criterion $|z_{1}-z_{2}| \leq \sigma_z$, where $z_{1} = \lambda_{c,1}/\lambda_1-1$, $z_{2} = \lambda_{c,2}/\lambda_2-1$, where $\lambda_1$ and $\lambda_2$ are the respective rest-frame wavelengths of the $\lambda 2796$ and $\lambda 2803$ transitions,
and where the variance in the quantity $|z_{1}-z_{2}|$ is
\begin{equation}
    \sigma_z^2 
    = \left[ \frac{\sigma_{\lambda_{c,1}}}{\lambda_1} \right]^2 + 
    \left[ \frac{\sigma_{\lambda_{c,2}}}{\lambda_2} \right]^2 \, .
\end{equation}
Second, we test if the doublet ratio, ${\it DR} = W_{1}/W_{2}$, is consistent with the physically allowed range using the condition 
\begin{equation}
(1-\sigma_{\it DR}) \leq {\it DR} \leq (2 + \sigma_{\it DR}) \, ,
\end{equation}
where 
\begin{equation}
\sigma^2_{\it DR} = {\it DR}^2 
\left(
\left[ \frac{\sigma^{}_{W_1}}{W_1} \right]^2
+
\left[ \frac{\sigma^{}_{W_2}}{W_2} \right]^2
\right) \,
\end{equation}
is the uncertainty in ${\it DR}$, which, on average is 0.24.  The average uncertainty in If both conditions are satisfied, we consider the two lines to be a candidate doublet and save the absorption properties for further vetting and processing. We simply define the absorption redshift as $z_{\rm abs} = z_1$.

If, during our search for the number of minimum in $w_j$ within the aperture from $j_-$ to $j_+$, we locate two minima, we deblended the absorption feature using two-component Gaussian decomposition. We then compute the equivalent widths and their uncertainties, and perform the two checks for a consistent redshift for the fitted line centers and a physically consistent doublet ratio.  If both conditions are satisfied, we assign $z_{\rm abs} = z_1$ and save the absorption properties for further vetting and processing. In the left panels of Figure~\ref{fig:flowchart}, we present a schematic flowchart of the detection algorithm and candidate doublet testing.

\subsection{Completeness of the Detection Algorithm} \label{sec: Detection completeness}

We assessed our detection algorithm using a Monte Carlo simulation to compute the completeness function 
\begin{equation}
    C(W_r, {\it DR}, z) = \frac{n_{\rm det}}{n} \, ,
\label{eq:ourcompleteness}
\end{equation}
which we define as the fraction of systems we successfully detect as a function of ${\MgII}$~$\lambda 2796$ rest-frame equivalent width, {\MgIIdblt} doublet ratio, and redshift.  In Eq.~\ref{eq:ourcompleteness},  $n$ is the total number of known doublets with ${\MgII}$~$\lambda 2796$ rest-frame equivalent width $W_r$, doublet ratio {\it DR}, and  redshift $z$, and $n_{\rm det}$ is the number of those detected.

In the right-hand panel of Figure~\ref{fig:flowchart}, we present a schematic of the steps to compute the completeness function. First, we ``clean" the OzDES spectra of all  absorption lines redward of the {\Lya} emission line. That is, the flux values in pixels associated with absorption features objectively located at the  $N_\sigma=3$ level (as described in Section~\ref{sec:locatingdoublets}) were replaced with the fitted continuum flux. These replacement continuum values were given simulated Gaussian noise characteristics matching those of the continuum local to the replaced pixels. We then ran our detection algorithm on all cleaned quasar spectra to verify that no absorption lines were detected at the $N_\sigma = 3$ level.

\begin{figure}[ht] 
\centering
\includegraphics[width=0.98\columnwidth]{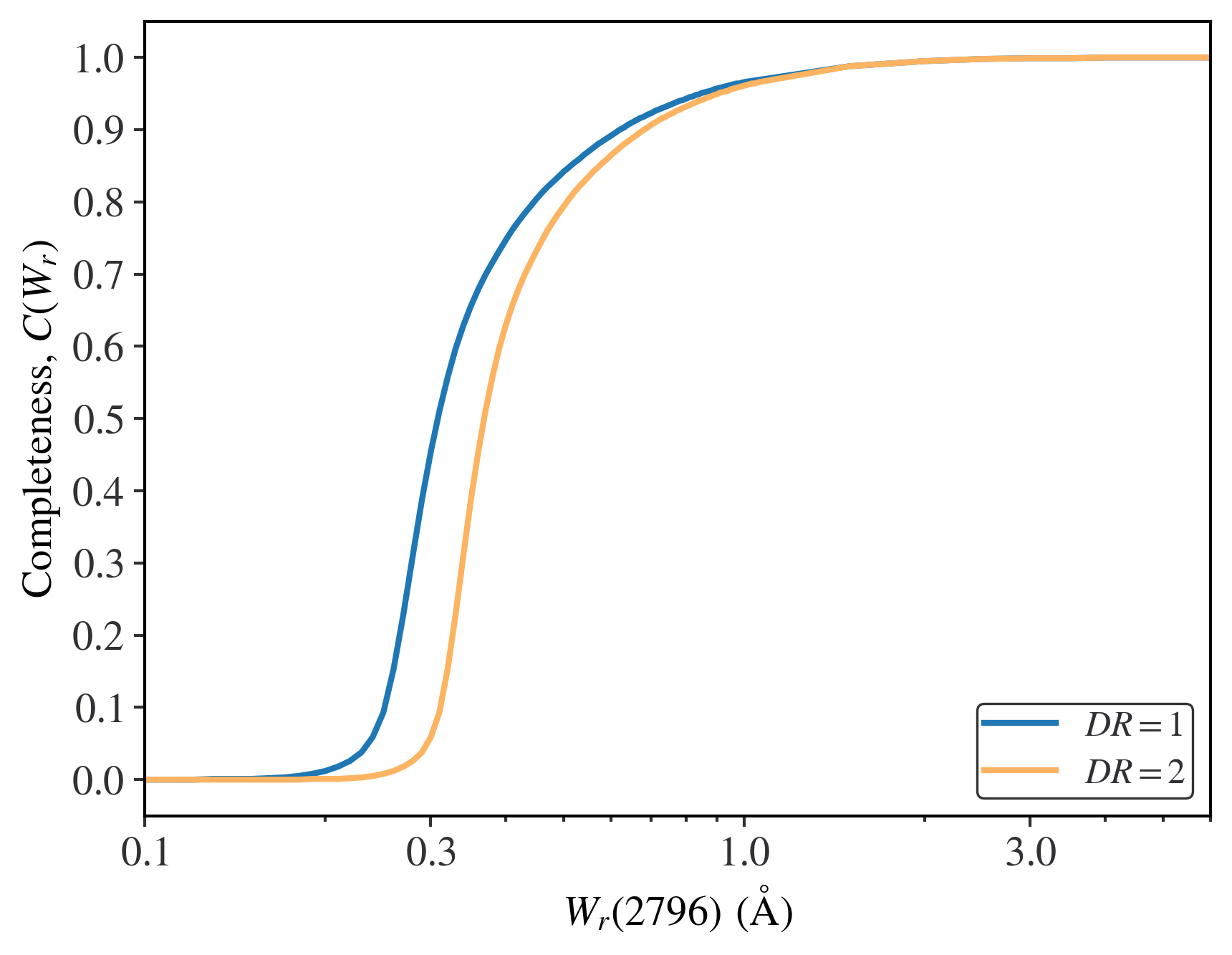}
\captionsetup{justification=justified, singlelinecheck=off} 
\caption{The completeness function $C(W_r, z)$ for $0.3 \leq z \leq 1.26$, for the {\MgII} doublet detection algorithm describe in Section~\ref{sec: Detection completeness}. The blue curve is for ${\it DR} = 1$ absorbers.  The orange curve is for ${\it DR} = 2$ absorbers. The completeness is $\sim 50$\% for $W_r = 0.3$~{\AA}.}
\label{fig:Det Algo Completeness}
\end{figure}

\begin{deluxetable*}{ccccc}
\tablewidth{0pt}
\tablecaption{OzDES {\MgII} Absorber catalog\label{table:OzDES catalog}}
\tablehead{
\colhead{OzDES Quasar ID\tablenotemark{$\dagger$}} & 
\colhead{$z_{\rm abs}$} &
\colhead{Wavelength ({\AA})} &
\colhead{$W_r(2796)$~({\AA})} &
\colhead{$W_r(2803)$~({\AA})}
}
\startdata
2925344837 & 0.602 & 4480.02 & 1.12$\pm$0.12 & 0.72$\pm$0.11 \\
2925352689 & 1.228 & 6229.65 & 0.38$\pm$0.10 & 0.43$\pm$0.10 \\[-4pt]
 & 1.663 & 7447.96 & 0.35$\pm$0.06 & 0.19$\pm$0.06 \\
2925358447 & 0.756 & 4911.72 & 1.14$\pm$0.14 & 0.70$\pm$0.12 \\[-4pt]
 & 0.840 & 5145.21 & 1.48$\pm$0.13 & 0.89$\pm$0.12 \\[-4pt]
 & 0.889 & 5282.71 & 1.02$\pm$0.11 & 0.81$\pm$0.11 \\
2925360137 & 0.759 & 4918.47 & 1.61$\pm$0.40 & 1.05$\pm$0.33 \\[-4pt]
 & 0.937 & 5416.58 & 0.78$\pm$0.21 & 0.63$\pm$0.17 \\[-4pt]
 & 1.224 & 6217.72 & 0.41$\pm$0.13 & 0.60$\pm$0.18 \\
2925362319 & 1.652 & 7414.75 & 0.31$\pm$0.06 & 0.22$\pm$0.05 \\
\enddata 
\vglue 0.1in
\tablenotetext{\dagger}{OzDES identifiers have the prefix SVA1-COADD}
\tablecomments{Table~\ref{table:OzDES catalog} is published in its entirety in machine-readable format. A portion is shown here for guidance regarding its form and content.}
\end{deluxetable*}

To construct the completeness function, we defined a grid of redshift and equivalent width bins. We chose a bin width of $\Delta z = 0.1$ covering the domain of the survey redshift, $z \in (0.3,2.2)$.  We also chose $W_r$ bin widths of $\Delta W_r = 0.1$~{\AA} over the domain $W_r \in (0.1,4.0)$~{\AA}. For a given cleaned quasar spectrum, the completeness function is built up by stepping through the redshift and equivalent width bins. For a given $(W_r,z)$ bin, we generated a mock {\MgII} doublet pair using a double Gaussian function (see Eq.~\ref{eqn: Double Gaussian}).  First, we generated a uniform random deviate $A_1 \in (0,1)$, which represents the flux decrement in the center of the $\lambda 2796$ line.  Using Eq.~\ref{eq:Gauss-EW}, we computed the Gaussian line width, $\Sigma = W_1/(\sqrt{2\pi}A_1)$, where $W_1=W_r(1+z)$ is the observed wavelength of the $\lambda 2796$ line.  If $\Sigma < \sigma_j$, where $\sigma_j$ is the ISF Gaussian standard deviation given by Eq.~\ref{eq:LSFwidth} evaluated at $\lambda_j = 2796.352(1+z)$, we set  $\Sigma = \sigma_j$. This ensured that no mock lines are narrower than the instrumental resolution.  We assumed both lines of the doublet experience the same line broadening, and therefore have the same $\Sigma$. We then generated a random deviate ${\it DR} \in (1,2)$, and determined the flux decrement of the $\lambda 2803$ line, $A_2=A_1/{\it DR}$.  This yields $W_2 = \sqrt{2\pi}A_2\Sigma = \sqrt{2\pi}A_1\Sigma/{\it DR} = W_1/{\it DR}$.

Given $A_1$, $A_2$, $\Sigma$, $\lambda_{c,1} = 2796.352(1+z)$, and $\lambda_{c,2} = 2803.531(1+z)$, we used Eq.~\ref{eqn: Double Gaussian} to generate a {\MgII} absorption doublet and inserted it into the cleaned spectrum. We then ran our detection algorithm and implement our doublet candidate criteria tests. This was done blind to the known redshift of the mock doublet. To account for the range of doublet ratios, for each $(W_r,z)$ bin we generated, inserted, and tested 10 doublets with random ${\it DR} \in (1,2)$.  After all $(W_r,z)$ bins were implemented for a given cleaned quasar, we stepped to the next cleaned quasar in the sample until all quasars were included.  We then computed $C(W_r,{\it DR},z)$ from Eq.~\ref{eq:ourcompleteness}.

The expectation is that a higher percentage of doublets with ${\it DR} \simeq 1$ will be detected compared to those with ${\it DR} \simeq 2$ because of the stronger absorption in the $\lambda 2803$ line.  To examine this quantitatively, we undertook the calculation of $C(W_r,1,z)$, holding ${\it DR} = 1$, and the calculation of $C(W_r,2,z)$, holding ${\it DR} = 2$.
In Figure~\ref{fig:Det Algo Completeness}, we show the detection completeness for ${\it DR} = 1$ and ${\it DR} = 2$ over the redshift range $0.3 \leq z \leq 2.2$.  There is a clear drop in the detection completeness for small equivalent widths, with 50\% completeness occurring at $W_r = 0.3$~{\AA}.

\subsection{The Absorber Sample}

\begin{figure}[ht] 
\centering
\includegraphics[width=0.98\columnwidth]{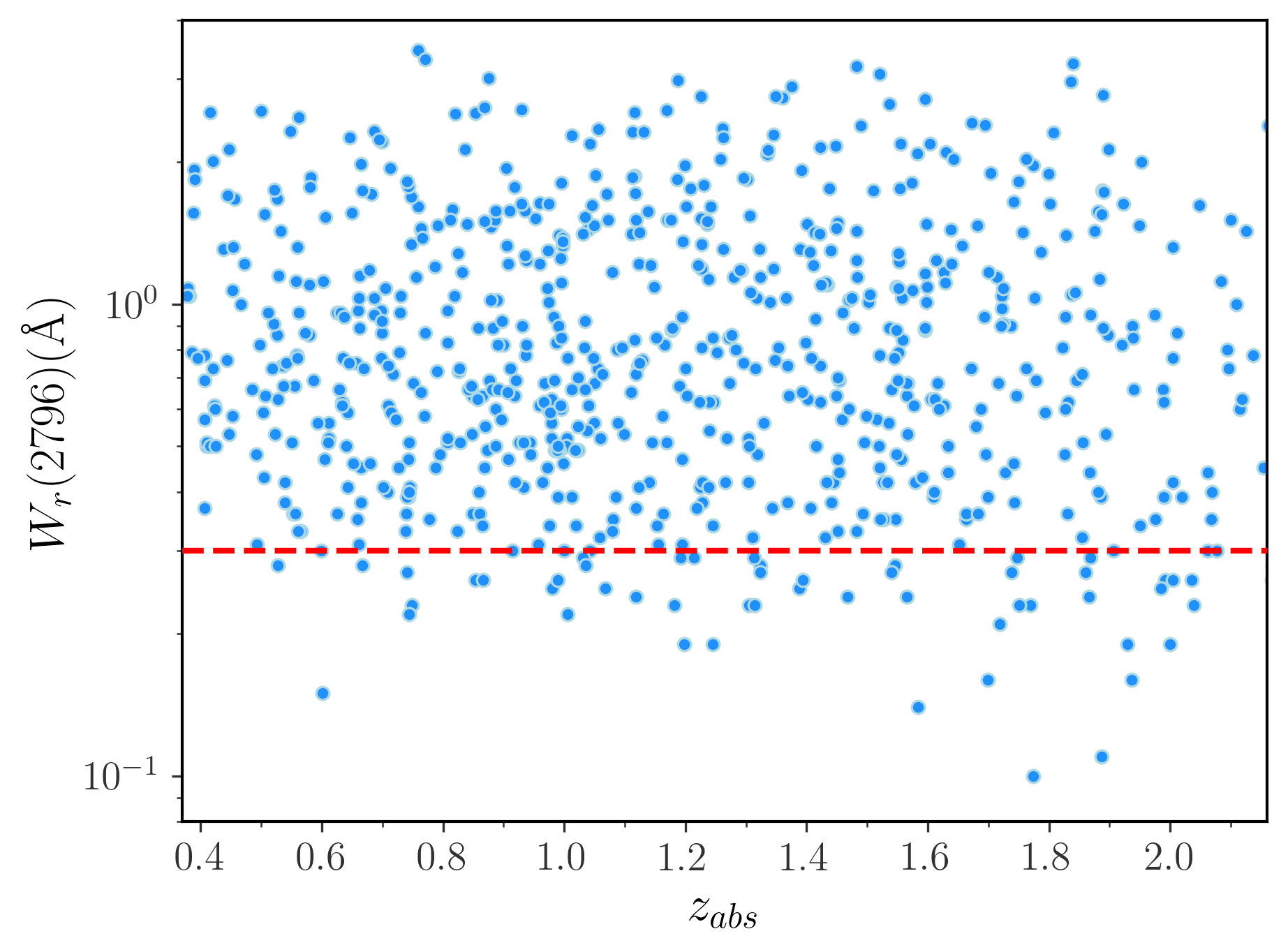}
\captionsetup{justification=justified, singlelinecheck=off} 
\caption{
The rest-frame equivalent width of the {\MgII}~$\lambda 2796$ transition versus redshift for the full OzDES {\MgII} absorber catalog of 717 absorbers. The red dashed horizontal line is $W_r = 0.3$~{\AA}, corresponds to 50\% completeness. Absorbers above line comprise our science sample of 656 absorbers.}
\label{fig:EW vs z}
\end{figure}

We found 717 {\MgII} doublets over the redshift range $0.33 \leq z \leq 2.19$ with equivalent widths in the range $0.1 \leq W_r \leq 3.45$~{\AA}. 
In Table~\ref{table:OzDES catalog} we present a partial list of the sample. In Figure~\ref{fig:EW vs z}, we show the distribution of rest-frame equivalent widths of the {\MgII}~$\lambda 2796$ transition versus redshift for the full catalog.

\begin{figure*}[th]
\centering
\includegraphics[width=0.93\textwidth]{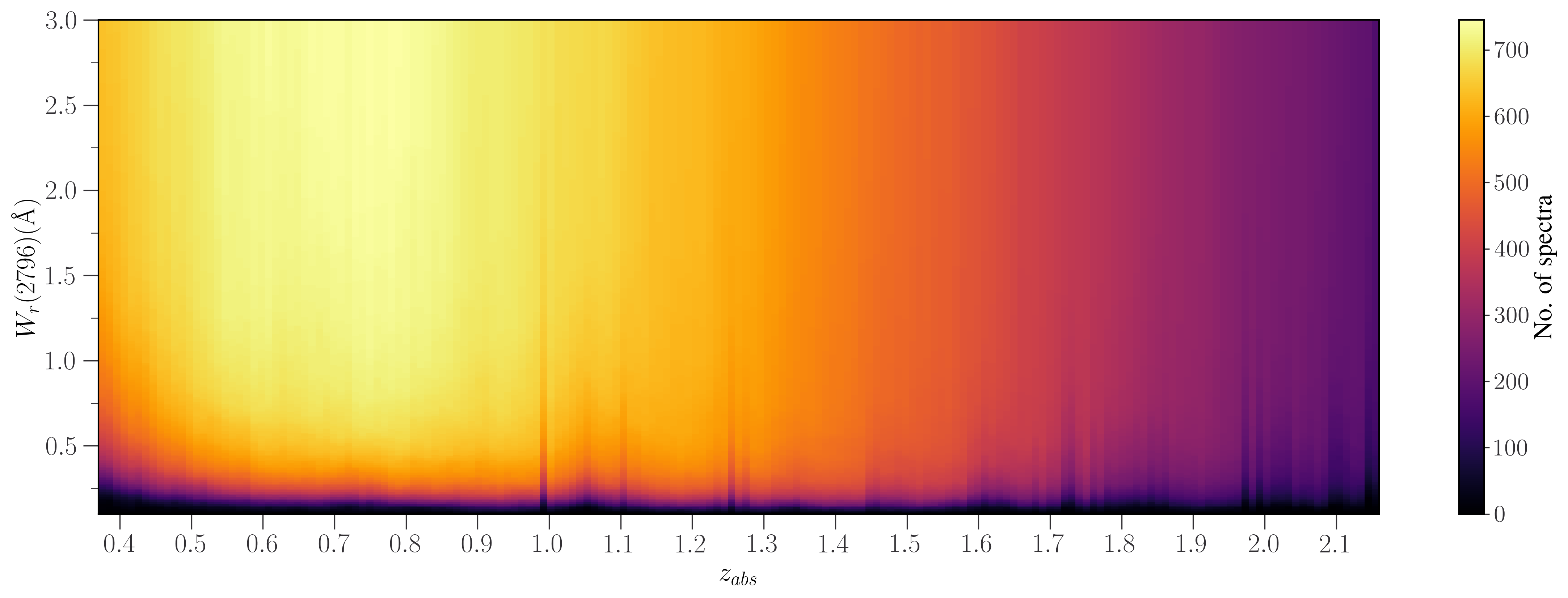}
\captionsetup{justification=justified, singlelinecheck=off} 
\caption{Redshift path sensitivity function g(z,$W_{min}$) for the {\MgII} 2796\AA at the 3$\sigma$ significance level i.e. $N_\sigma = 5$. The plot shows the number of quasar spectra in which a {\MgII} absorption line of rest frame equivalent width $W_r$ could have been detected, as a function of redshift. The dark vertical bands signify bright night-sky lines or atmospheric absorption. The depression at $z\approx1.05$ corresponds to the wavelength where the dichroic splits the two arms of the spectrograph.}
\label{fig:EWS}
\end{figure*}

The detection completeness is 50\% at $W_r = 0.3$.  We adopt this minimum equivalent width threshold or our science sample.  The science sample comprises all system above the horizontal dashed red line in Figure~\ref{fig:EW vs z}. The science sample is has  
656 systems over the redshift range $0.33 \leq z \leq 2.19$ with equivalent widths in the range $0.3 \leq W_r(2796) \leq 3.45$~{\AA}.

\subsection{OzDES Redshift Path}

The redshift path sensitivity function, $g(W,z)$, 
provides the number of quasar spectra in the survey in which which a {\MgII} absorption system with rest-frame equivalent width $W_r$ or greater could be detected at redshift $z$. Following \citet{Lanzetta_87} and \citet{steidel1992}, this function is computed as sum over all $N$ quasars in the sample
\begin{equation} 
    g(W_r,z) = \sum_{q=1}^{N} H(\Delta z_l)H(\Delta z_u)H(W_r - N_\sigma \sigma_{w_r}) \, ,
\label{eqn:g(W,z)}
\end{equation}
where $H(x)$ is the Heaviside function, $\Delta z_l = z \!-\! z^{min}_q$ and $\Delta z_u = z^{max}_q \!-\! z$, where $z^{max}_q$ and $z^{min}_q$ refer to the maximum and minimum redshifts adopted for the $q$th quasar spectrum, and where $\sigma_{w_r} = \sigma_{w}/(1+z)$ is the rest-frame limiting equivalent width detection threshold at redshift $z$ in spectrum $q$, where $\sigma_w$ is given by Eq.~\ref{eq:schneider-wsig}. We adopted $N_{\sigma}=5$.  We avoid searching through the {\Lya} forest by adopting $z^{min}_q = 1215.67(1+z_q)$ when the quasar {\Lya} emission line falls redward of 4200~{\AA}, corresponding to the lower limit of our survey.  We also limit our survey to $cz^{max}_q = -5000$~{\kms} from the {\MgII} emission of the quasar \citep[see, for example,][]{weymann_1981_absorption}.

In Figure~\ref{fig:EWS}, we present a ``heatmap'' showing the number of quasar spectra in which a {\MgII} absorber at a redshift $z$ with a rest-frame equivalent width $W_r$ ot greater could have been detected. The spectrograph used for the survey, AAOmega, consists of red and blue arms that are split by a dichroic. This leads to a discontinuity in the spectra that is reflected in the plot at $5700$~{\AA} at $z \simeq 1.0$ \citep{childress_17}. The dark vertical lines in the figure show atmospheric absorption bands.

Using our definition of $g(W,z)$, we can define the comoving path $\Delta X$. The comoving path $\Delta X$ is a measure of the distance that the light from a quasar has traveled to reach the observer. This is an important quantity in the context of an absorption line survey, because it represents the length of the path that the survey covers in redshift space. For example, a survey with a large comoving path length covers a larger volume of the Universe and is therefore more sensitive to rare or distant absorbers. $\Delta X$ can be thought of as a measure of the survey depth or volume. Over a finite redshift interval from $z_1$ to $z_2$, this is expressed analytically as
\begin{equation} \label{eqn:Delta X}
    \Delta X(W_r) = \int_{z_1}^{z_2} g(W_r, z)\frac{dX}{dz}dz \,,
\end{equation}
where, assuming a flat Universe,
\begin{equation}
    \frac{dX}{dz} = \frac{(1+z)^2}{\Omega_M(1+z)^3 + \Omega_{\Lambda}} \, .
\end{equation}
This term represents the comoving distance per unit redshift, which is a fundamental quantity that characterizes the expansion history of the Universe.

\section{Results} 
\label{sec:Results}

The main results we present are the {\MgII}~$\lambda 2796$ rest-frame equivalent width distribution and the redshift path density of {\MgII} doublets.   These comprise the two most extensively studied distributions measured for {\MgII} absorption systems in the northern hemisphere using low resolution quasar spectroscopy \citep[e.g.,][]{steidel1992, nestor_2005_mgiiabsorption, Seyffert_13, zhu_2013_the}.  Thus, they provide robust data by which to compare our first view of {\MgII} absorber characteristics in the southern hemisphere.

In Section~\ref{sec:massdensity}, we introduce a method for estimating the cosmic mass density of Mg$^{+}$ ions using low-resolution surveys as informed by high-resolutions survey line measurements.
 
\subsection{Equivalent Width Distribution}
\label{sec:ewdistribution}

The equivalent width distribution is defined as the probability density function $f(W_r) \!=\! d^2N/dW_r dX$.  This gives the relative number of absorption systems with rest-frame equivalent width $W_r$ in redshift range $z$ per unit equivalent width per unit comoving redshift path.

\begin{figure}[ht]
\centering
\includegraphics[width=0.98\columnwidth]{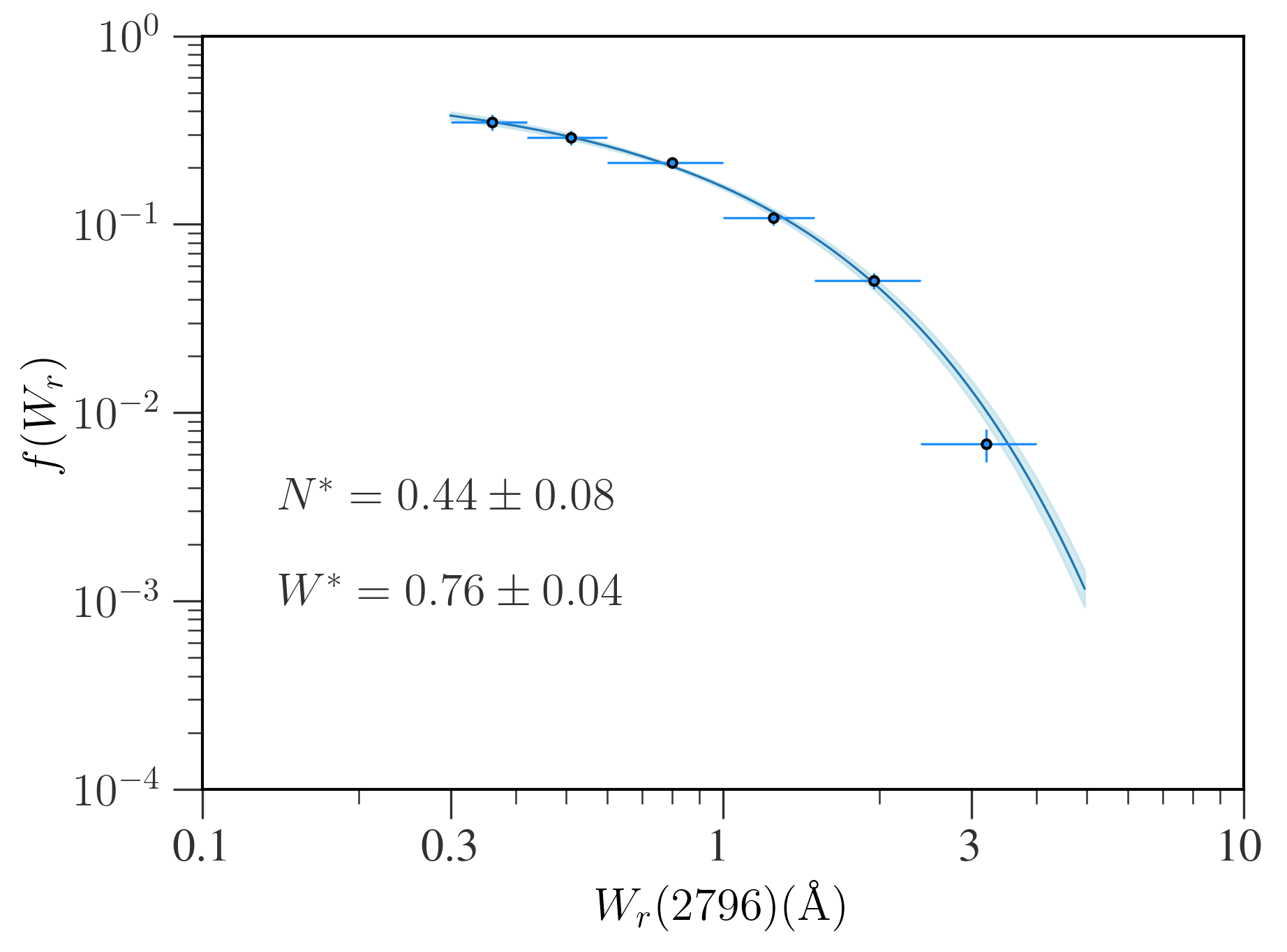}
\captionsetup{justification=justified, singlelinecheck=off} 
\caption{The rest-frame equivalent width distribution function, $f(W_r) = d^2W_r/dW_r dX$, of the {\MgII}~$\lambda 2796$ transition for the OzDES science sample. The vertical error bars are the Poisson uncertainty. The horizontal bars are the equivalent width bin sizes.  The solid blue curve is the fit for the exponential function (Eq.~\ref{eqn:EW Dist fit}). Comparison of the fitted parameters $W_*$ and $N_*$ with other surveys are listed in Tale~\ref{table:W* comparison}.}
\label{fig:n(W_r)}
\end{figure}

In practice, the equivalent width distribution is computed in discrete bins. For a given equivalent width bin $\Delta W_r$ centered at $W_r$ over a given comoving redshift path $\Delta X(W_r)$, as defined in Eq.~\ref{eqn:Delta X}, we have 
\begin{equation} \label{eqn:EW Dist}
    f(W_r) = \frac{N}{\Delta W_r \Delta X(W_r)} \,,
\end{equation}
where $N$ is the number of absorbers in the equivalent bin. 
In Figure~\ref{fig:n(W_r)}, we show the equivalent width distribution of {\MgII} $\lambda 2796$~{\AA} absorbers in our sample. The points are the mean $W_r$ in the bins and the horizontal error bars are the equivalent width bin sizes.  The vertical error bars are based on Poisson statistics. 

For surveys sensitive to $W_r < 0.05$~{\AA}, a Schechter function \citep{schechter76} provides a good description of the the distribution function \citep{Kacprzak_Schechter, mathes2017}. However, since we used low resolution spectra for this work, we are 50\% complete at $W_r \simeq 0.3$~{\AA}.  For such completeness levels, the distribution is dominated by the exponential term \citep[e.g.,][]{steidel1992, nestor_2005_mgiiabsorption, cooksey2013, Seyffert_13, zhu_2013_the}. We thus adopt
\begin{equation} 
    f(W_r) = \frac{N_{*}}{W_{*}} 
    \exp \left( -\frac{W}{W_{*}}\right) \,.
\label{eqn:EW Dist fit}
\end{equation}
Using the LMFIT python package \citep{lmfit}, we obtained best-fitted parameters of $N_{*} = 0.44\pm0.08$ and $W_{*} = 0.76\pm 0.04$~{\AA}. We preformed the fit on the binned data. The bins have equal widths in dex.
 
In Figure~\ref{fig:n(W_r)}, the solid blue line represents the best-fit exponential function. 
The shape of the equivalent width distribution function aligns with findings from earlier low-resolution surveys \citep[e.g.,][]{steidel1992, nestor_2005_mgiiabsorption, Seyffert_13, zhu_2013_the}. A comparison of our $f(W_r)$ fit parameters $N_*$ and $W_*$ to those from \citet{steidel1992}, \citet{nestor_2005_mgiiabsorption}, and \citet{Seyffert_13} is presented in Table~\ref{table:W* comparison}. It is essential to note that neither \citet{steidel1992} nor \citet{nestor_2005_mgiiabsorption} normalize their equivalent width distribution by the comoving redshift path $\Delta X(W_r)$. Consequently, our normalization $N_*$ is directly comparable only with \citet{Seyffert_13}, who fitted exponential function using the $f(W_r) = k\exp(\alpha W_r)$. To equate Eq.~\ref{eqn:EW Dist fit}, we converted $k$ and $\alpha$ from \citet{Seyffert_13} to $N_*$ and $W_*$ using the relations: $W_* = -1/\alpha$ and $N_* = kW_*$.  

\begin{deluxetable}{ccc}
\tablewidth{0pt}
\tablecaption{Equivalent Width Fitted Parameters\label{table:W* comparison}}
\tablehead{
\colhead{Reference} & 
\colhead{$W_*$~({\AA})} &
\colhead{$N_*$} 
}
\startdata
\citet{steidel1992}\tablenotemark{$\dagger$} & 0.66$\pm$0.11 & 1.55$\pm$0.20 \\
\citet{nestor_2005_mgiiabsorption}\tablenotemark{$\dagger$} & 0.702$\pm$0.017 & 1.187$\pm$0.052 \\
\citet{Seyffert_13} & 0.714$\pm$0.005 & 0.506 $\pm$ 0.015 \\
OzDES (this work) & 0.76 $\pm$ 0.04 & 0.44$\pm$ 0.08 \\
\enddata 
\tablenotetext{\dagger}{Normalization is $dN/dz$. $N^*$ not comparable to OzDES.}
\end{deluxetable}

\begin{figure*}[ht]
\centering
\includegraphics[width = 0.75\textwidth]{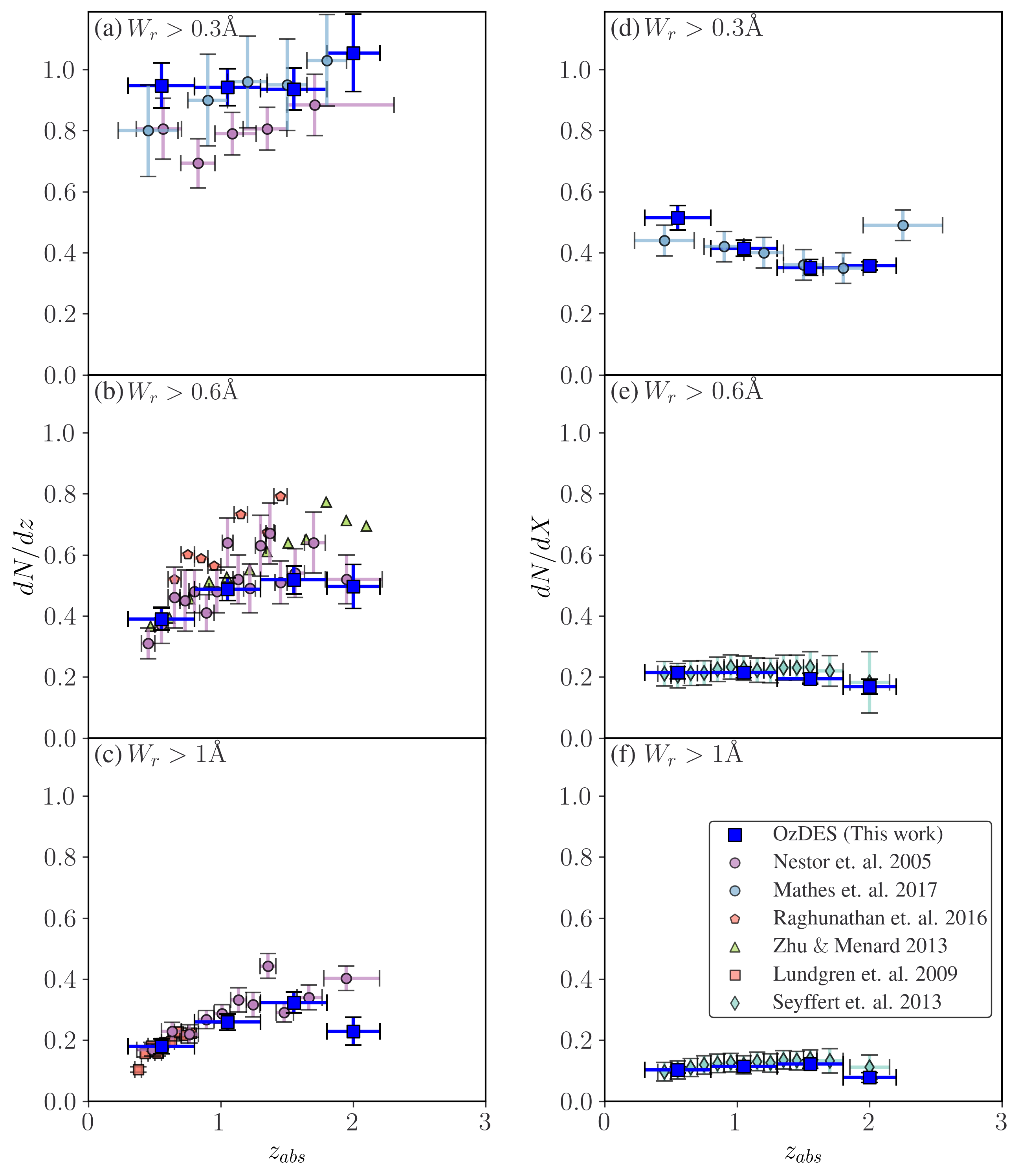}
\captionsetup{justification=justified, singlelinecheck=off} 
\caption{(left panels) The redshift path density $d{\cal N}/dz$ of {\MgII} absorber samples with lower threshold of (a) $W_r \geq 0.3$~{\AA}, (b) $W_r \geq 0.6$~{\AA}, and (c) $W_r \geq 1$~{\AA}.  (right panels) The comoving path density $d{\cal N}/dX$ for (d) $W_r \geq 0.3$~{\AA}, (e) $W_r \geq 0.6$~{\AA}, and (f) $W_r \geq 1$~{\AA}.  The OzDES systems are the blue data to which we compare the available northern hemisphere works of \citet{steidel1992}, \citet{nestor_2005_mgiiabsorption},  \citet{Seyffert_13}, \citet{zhu_2013_the}, \citet{Raghunathan2016}, and \citet{mathes2017}. Note, that in panel (b) \citet{Raghunathan2016} compute $d{\cal N}/dz$ for $W_r>0.65$~{\AA}.}
\label{fig:dNdz}
\end{figure*}

\begin{deluxetable*}{ccccccccc}
\tablewidth{0pt}
\tablecaption{$d{\cal N}/dz$, $d{\cal N}/dX$ and $\Omega\subMgII \times 10^{-6}$\label{tab:omegamg2}}
\tablehead{
\colhead{} & 
\multicolumn{2}{c}{($W_r \geq 0.3$~{\AA})} &
\multicolumn{2}{c}{($W_r \geq 0.6$~{\AA})} &
\multicolumn{2}{c}{($W_r \geq 1.0$~{\AA})} &
\colhead{(Sum)} &
\colhead{(Integral)}\\[-6pt]
\colhead{Redshift} & 
\colhead{$d{\cal N}/dz$} &
\colhead{$d{\cal N}/dX$} &
\colhead{$d{\cal N}/dz$} &
\colhead{$d{\cal N}/dX$} &
\colhead{$d{\cal N}/dz$} &
\colhead{$d{\cal N}/dX$} &
\colhead{$\Omega\subMgII(z)$} &
\colhead{$\Omega\subMgII(z)$}  
}
\startdata \\[-10pt]
(0.33,0.80] & $0.95\pm 0.07$ & $0.51\pm 0.04$ & $0.39\pm 0.04$ & $0.21\pm 0.02$ & $0.18\pm 0.02$ & $0.10\pm 0.01$ & $0.47\substack{+0.26 \\ -0.11}$ & $0.76\substack{+0.28 \\ -0.16}$\\
(0.80,1.30] & $0.94\pm 0.06$ & $0.41\pm 0.03$ & $0.49\pm 0.04$ & $0.21\pm 0.02$ & $0.26\pm 0.03$ & $0.11\pm 0.01$ & $0.42\substack{+0.18 \\ -0.08}$ & $0.69\substack{+0.18 \\ -0.13}$\\
(1.30,1.80] & $0.93\pm 0.06$ & $0.35\pm 0.03$ & $0.52\pm 0.04$ & $0.19\pm 0.02$ & $0.32\pm 0.03$ & $0.12\pm 0.01$ & $0.41\substack{+0.16 \\ -0.08}$ & $0.65\substack{+0.19 \\ -0.14}$\\
(1.80,2.20) & $1.05\pm 0.12$ & $0.36\pm 0.01$ & $0.50\pm 0.07$ & $0.17\pm 0.02$ & $0.23\pm 0.04$ & $0.08\pm 0.02$ & $0.24\substack{+0.26 \\ -0.08}$ & $0.47\substack{+0.32 \\ -0.11}$\\[2pt]
\enddata
\end{deluxetable*}

Comparing $W_*$ values obtained by \citet{steidel1992}, \citet{nestor_2005_mgiiabsorption}, and \citet{Seyffert_13}, we find that the OzDES value is statistically in agreement, if not trending a slightly higher . As shown by \citet[][see their Figure~9]{chen-simcoe2017}, the value of $W_*$ is known to evolve over the redshift range of the OzDES sample, $0.3\leq z\leq 2.2$. The value decreases from $W_* \simeq 0.9$~{\AA} at $z\simeq 2.2$ to $W_* \simeq 0.6$~{\AA} at $z\simeq 0.3$.  All $W_*$ values listed in Table~\ref{table:W* comparison} are determined over virtually identical redshift ranges for which the median redshifts of the samples are $\langle z \rangle \sim 1.1$. The median redshift of the OzDES science sample is $\langle z\rangle=1.14$.  We note that the OzDES measurement of $W_* = 0.76$ is fully consistent with the maximum likelihood value of $W_*= 0.74$~{\AA} found by \citet[][see their Table~5]{chen-simcoe2017} for this redshift based on a meta-study of {\MgII} surveys covering $0.37 \leq z \leq 7.1$.  Comparing the OzDES $N_*$ value with \citet{Seyffert_13}, OzDES is statistically consistent. If we renormalize the OzDES $N_*$ to yield the equivalent width per unit redshift, i.e., $f(W_r)\cdot \{ \Delta X(W_r)/ \Delta Z(W_r) \}$, we can compare the $N_*$ values of \citet{steidel1992} and \citet{nestor_2005_mgiiabsorption}.  For OzDES, $\Delta Z(0.3) = 690$, and $\Delta X(0.3) = 1614$, giving $\Delta Z(0.3)/ \Delta Z(0.3) = 2.34$, which yields $N_* = 1.02 \pm 0.19$ for OzDES.  This value is consistent with that of \citet[][$N_* = 1.19 \pm 0.05$]{nestor_2005_mgiiabsorption}, but smaller than that value measured by \citet{steidel1992}.

\subsection{Absorber Path Density}

The redshift path density, $d{\cal N}/dz$, quantifies the number of absorbers per unit redshift for absorbers above a given $W_r$ threshold.  For a survey in which the $W_r$ threshold is variable with redshift, $d{\cal N}/dz$ is defined as \citep[e.g.,][]{Lanzetta_87},
\begin{equation}
\begin{array}{rcl}
   \displaystyle \frac{d{\cal N}}{dz} &=& \displaystyle\sum_i \frac{1}{\Delta Z(W_i)} \\[15pt]
    \displaystyle \sigma_{d{\cal N}/dz} &=& \displaystyle\sum_i \frac{1}{\Delta Z^2(W_i)} \, ,
\end{array}
\end{equation}
where $\Delta Z(W_i)$ is the total redshift path covered in the survey for absorber $i$ having rest-frame equivalent width $W_i$. $\sigma_{d{\cal N}/dz}$ is the uncertainty in $d{\cal N}/dz$. The sum is taken over all absorbers in the redshift interval for which $d{\cal N}/dz$ is being measured.  If the redshift interval spans from $z_1$ to $z_2$, then the redshift path for a system with $W_r=W_i$ in this interval is given by
\begin{equation}
    \Delta Z(W_r) = \int_{z_1}^{z_2} \!\!\! g(W_r,z) dz \,.
\end{equation}
Similarly, $d{\cal N}/dX$, the number of absorbers per unit comoving path length, is defined as
\begin{equation}
\begin{array}{rcl}
    \displaystyle \frac{d{\cal N}}{dX} &=& \displaystyle\sum_i \frac{1}{\Delta X(W_i)} \\[15pt]
    \displaystyle \sigma_{d{\cal N}/dX} &=& \displaystyle\sum_i \frac{1}{\Delta X^2(W_i)} \, ,
\end{array}
\end{equation}
where $\Delta X(W_i)$ is given by Eq.~\ref{eqn:Delta X} for absorber $i$ having rest-frame equivalent width $W_i$.

In Figure~\ref{fig:dNdz}, we show $d{\cal N}/dz$ and $d{\cal N}/dX$ as a function of redshift for {\MgII} absorbers in the OzDES survey (blue data points).  These data are listed in Table~\ref{tab:omegamg2}. We examine the path densities in four redshift bins such that they each roughly have 150 absorbers. We adopted and show subsamples having lower thresholds of $W_r \geq 0.3$~{\AA}, $W_r \geq 0.6$~{\AA}, and $W_r \geq 1$~{\AA} in order to directly compare our southern hemisphere measurements with 
the northern hemisphere data of \citet{steidel1992}, \citet{nestor_2005_mgiiabsorption}, \citet{lundgren2009}, \citet{Seyffert_13}, \citet{zhu_2013_the}, and \citet{Raghunathan2016}. 
For all subsamples, we find that the southern hemisphere path densities are consistent with the northern hemisphere path densities over the similar redshift ranges.

Not all northern surveys published results for all three equivalent width thresholds.  
Compared to \citet{nestor_2005_mgiiabsorption}, the OzDES $d{\cal N}/dz$ values for the $W_r \geq 0.3$~{\AA} sample are slightly higher, but appear to be in excellent agreement with \citet{steidel1992}.  For $W_r \geq 0.6$~{\AA}, the OzDES $d{\cal N}/dz$ values are consistent with the three surveys of  \citet{nestor_2005_mgiiabsorption}, \citet{zhu_2013_the}, \citet{Raghunathan2016}, though they appear to cluster toward the lower values within the ``cloud" formed by these data. For $W_r \geq 1.0$~{\AA}, the OzDES $d{\cal N}/dz$ values are consistent with the surveys of \citet{nestor_2005_mgiiabsorption} and \citet{lundgren2009}.  In terms of $d{\cal N}/dX$, the OzDES values are in full agreement with the published values of \citet{Seyffert_13} for $W_r \geq 0.6$~{\AA} and $W_r \geq 1$~{\AA}.  Though we do not show the comparison on Figure~\ref{fig:dNdz}, the OzDES $d{\cal N}/dX$ for the $W_r \geq 0.3$~{\AA} sample is consistent with the values from the high-resolution survey of \citet{mathes2017}.

\section{The Mass Density}
\label{sec:massdensity}

The mass density, $\Omega \!=\! \rho/\rho_c$, of an elemental species or ion is defined as the ratio of its cosmic mass density to the critical density of the universe, $\rho_c = 3H^2/8\pi G$.  
In general form, we can define the {\MgII} mass density as
\begin{equation} 
\Omega\subMgII (z)  = \frac{H_0}{c} \frac{m_{\hbox{\tiny Mg}}}{\rho_c} \langle N \rangle \frac{d{\cal N}}{dX}
\label{eq:Omega_MgII}
\end{equation}
where $\langle N \rangle$ is the measured mean column density of the population of {\MgII} absorbers, and $d{\cal N}/dX$ is their comoving path density.  We have
\begin{equation}
    C = \frac{H_0}{c} \frac{m_{\hbox{\tiny Mg}}}{\rho_c} = 3.214\times10^{-22}\, {\rm cm}^2 \, ,
\end{equation}
where $m_{\hbox{\tiny Mg}} = 4.036\times10^{-23}$~g is the mass of a magnesium atom and $\rho_c = 9.77\times10^{-30}$ g cm$^{-3}$ is the present-day critical density of the universe. 

The product $\langle N \rangle d{\cal N}/dX$ yields the total column density of the population per unit of comoving path length and can be written 
\begin{equation}
 \langle N \rangle \frac{d{\cal N}}{dX} = 
 \frac{N_{tot}}{\Delta X}  = 
  \frac{1}{\Delta X} \displaystyle \sum_{i} N_i \, ,
\label{eqn:Omega Sum}
\end{equation}
where the sum is taken over all absorbers in the sample over the finite redshift range for the calculation. Alternatively, if the column density distribution function, $f(N)$, is measured, we can write $\langle N \rangle d{\cal N}/dX$ in integral form as the first moment of the column density distribution function,
\begin{equation} 
  \langle N \rangle \frac{d{\cal N}}{dX} =  
  \langle N \rangle \int^{N\subU}_{N\subL} \!\! f(N)dN = 
  \int^{N\subU}_{N\subL} \!\! f(N)NdN
       \, ,
\label{eqn:Omega Integral}
\end{equation}
where $N\subL$ and $N\subU$ are the minimum and maximum observed column density of the sample, and the column density distribution function has been normalized to $d{\cal N}/dX$.

\begin{figure*}[ht]
\centering
\includegraphics[width=0.88\textwidth]{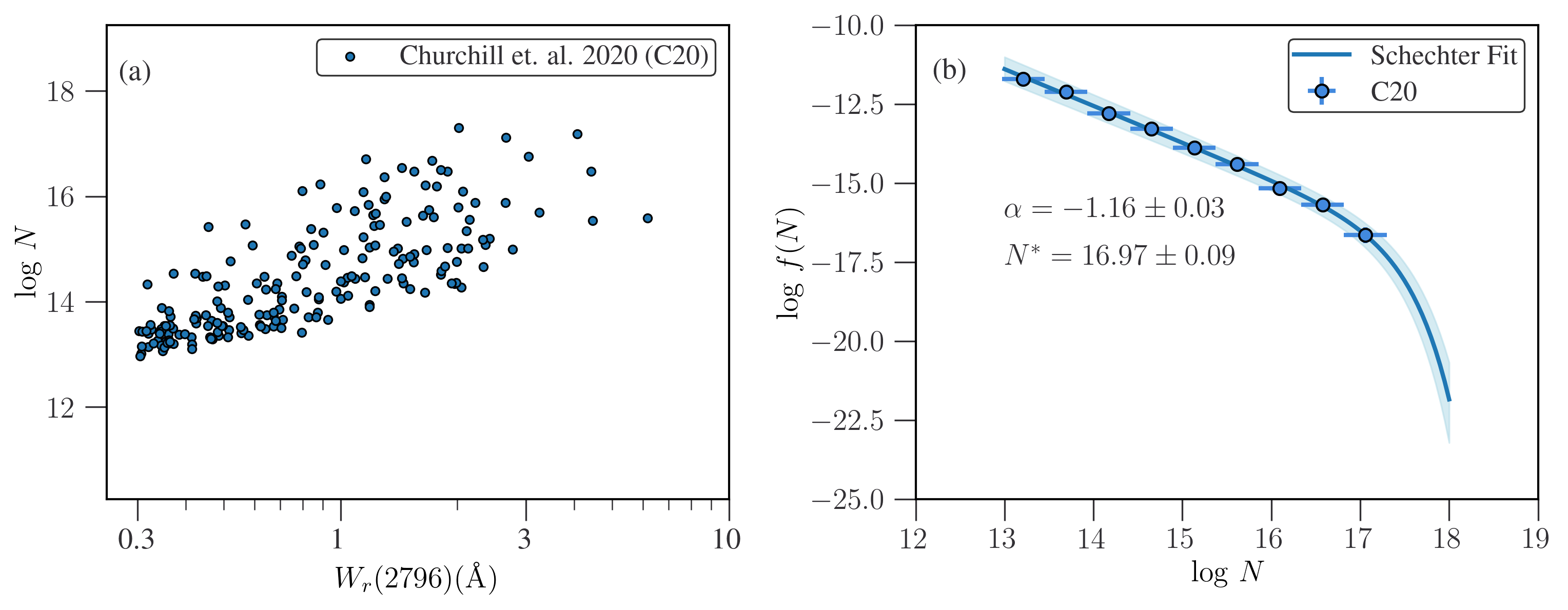}
\captionsetup{justification=justified, singlelinecheck=off} 
\caption{(a) The system-total column densities of {\MgII} absorbers obtained from Voigt profile fitting of high-resolution spectra from \citetalias{Churchill_2020} versus rest-frame equivalent width. Only absorbers with $W_r > 0.3$~{\AA} are included to ensure comparability with the OzDES science sample. (b) The column density distribution function of the \citetalias{Churchill_2020} sample.  The distribution is well fitted by a Schechter function (see Eq.~\ref{eq:thebigSchechter}) with $\alpha = -1.16\pm 0.03$ and $\log (N_*/{\rm cm}^{-2}) = 16.97 \pm 0.09$.}
\label{fig:log N vs log W}
\end{figure*}

Clearly, accurate estimation of the column densities of {\MgII} absorbers is crucial in determining the value of $\Omega\subMgII$. One of the most reliable methods for estimating column densities is to fit resolved absorption profiles with a Voigt profiles. Voigt profile fitting is appropriate only for high-resolution absorption line spectra where the atomic plus Gaussian line broadening can be resolved.  However, for our study, we only have low-resolution spectra available, and as a result, we must resort to alternative methods for estimating the column densities of the {\MgII} absorbers in our survey. 

Unfortunately, the relatively straight forward apparent optical depth method \citep[AOD, e.g.,][]{savage_aod91} is also not accurate for low resolutions spectra, as unresolved saturation can yield underestimates as large as $\sim\!1$~dex in the column densities \citep[also see][]{jenkins1996}.  Likewise, curve of growth (COG) methods also have their limitations. The equivalent widths of {\MgII} absorption lines reflect their multi-component nature in that they scale with the number of components, each of which has a unique column density \citep[e.g.,][]{petitjean1990, Churchill_2003, Churchill_2020}. A such, a unique COG relation cannot be applied to multi-component absorption lines.
 
\subsection{Monte Carlo Modeling} \label{sec:Column density estimation}

We have developed a Monte Carlo method for estimating the total column density ($N_{tot}$, Eq.~\ref{eqn:Omega Sum}) and the mean column density ($\langle N \rangle$, Eq.~\ref{eqn:Omega Integral}) for a survey of low-resolution absorption lines. The model is informed by the relationship between equivalent width and system-total column density from a database of high-resolution {\MgII} absorption profiles for which Voigt profile analysis has been performed. 

In summary, our approach to calculating $\Omega\subMgII$ is that we use a Monte Carlo method to obtain an estimated column density for each and every {\MgII} system in the OzDES survey.  This provides a single ``realization'' of the column densities for the entire observed sample of absorbers. We then compute $\Omega\subMgII(z)$ from Eq.~\ref{eq:Omega_MgII} using both the summation (Eq.~\ref{eqn:Omega Sum})
and integral (Eq.~\ref{eqn:Omega Integral}) formalism.  In this way, each Monte Carlo realization provides two estimates of the mass density.  We then implement $10^5$ Monte Carlo realizations of the OzDES sample from which we determine a mean mass density and its uncertainty.

In Figure~\ref{fig:log N vs log W}(a), we show the $N$--$W_r$ pairs from the sample of  \citet[][hereafter \citetalias{Churchill_2020}]{Churchill_2020}.  We use this sample to inform our Monte Carlo model.  We limited our study to systems with $W_r \geq 0.3$~{\AA}, as this equivalent width is representative of the minimum value in the OzDES science sample. We also limited our study to absorbers residing in the redshift range for detecting {\MgII} doublets in the OzDES spectra. A total of 198 systems met these selection criteria. In Figure~\ref{fig:log N vs log W}(b), we show the column density distribution function, which is well fitted by a Schechter function,
\begin{equation}
    f(N)dN = \Phi_0 \left( \frac{N}{N_*} \right)^{\alpha} \!\!
    \exp \left\{ -\frac{N}{N*} \right\}
    \frac{dN}{N_*}
\label{eq:thebigSchechter}
\end{equation}
where $\alpha$ is the power-law slope, $N_*$ is the characteristic column density, and $\Phi_0$ is the normalization. Using LMFIT \citep{lmfit}, we obtained least-square fitted parameters $\alpha=-1.16\pm0.03$  and $\log (N_*/{\rm cm}^{-2}) = 16.97\pm0.09 $.

Although the full \citetalias{Churchill_2020} sample can be convincingly fitted with a single Schechter function over the range $W_r \in (0.3,6.168)$~{\AA}, examination of Figure~\ref{fig:log N vs log W}(a) shows a clear trend between $N$ and $W_r$ indicating that the column density distribution function varies with equivalent width.  Note that there is a clear lower boundary to $N$ that increases with increasing $W_r$. In addition, for the smallest absorbers ($W_r \sim 0.3$~{\AA}) the range of $N$ is on the order of 1~dex, where as the range is $\sim\! 2$~dex for stronger absorbers ($W_r \sim 1.0$~{\AA}).  

In Appendix~\ref{sec:appA}, we detail how we treat the $W_r$ dependence of the column density distribution function and describe how the Monte Carlo model estimates the column density associated with a {\MgII} system of a given equivalent width. We also describe how we compute $N_{tot}$ and $\langle N \rangle$ for a given Monte Carlo realization of the OzDES survey.  In Appendix~\ref{sec:appB}, we quantify the degree to which the model is consistent with the column density distribution measured from the high-resolution {\MgII} sample. In Appendix~\ref{sec:appC}, we explain how we estimate $\Omega\subMgII$ and its uncertainty from $10^5$ Monte Carlo realizations of the OzDES science sample. 

\subsection{OzDES Mass Density}

In Table~\ref{tab:omegamg2}, we present $\Omega\subMgII$ for the science sample. We include both the summation method (Eq.~\ref{eqn:Omega Sum}) and the integral method (Eq.~\ref{eqn:Omega Integral}).  We computed $\Omega\subMgII$ in four redshift bins for $W_r \geq 0.3$~{\AA}.  We illustrate $\Omega\subMgII$ in Figure~\ref{fig:omegamg2} (also see Figure~\ref{fig:Omegadist} in Appendix~\ref{sec:appC}).  The red points represent the summation method and the purple points represent the integral method. In each redshift bin, we ran $10^5$ Monte Carlo realizations of the OzDES sample. The quoted best-values and their uncertainties are determined as described in Appendix~\ref{sec:appC}. Within the $1\sigma$ uncertainties, the summation and integral methods are consistent with one one another.

\begin{figure}[th] 
\centering
\includegraphics[width=0.98\columnwidth]{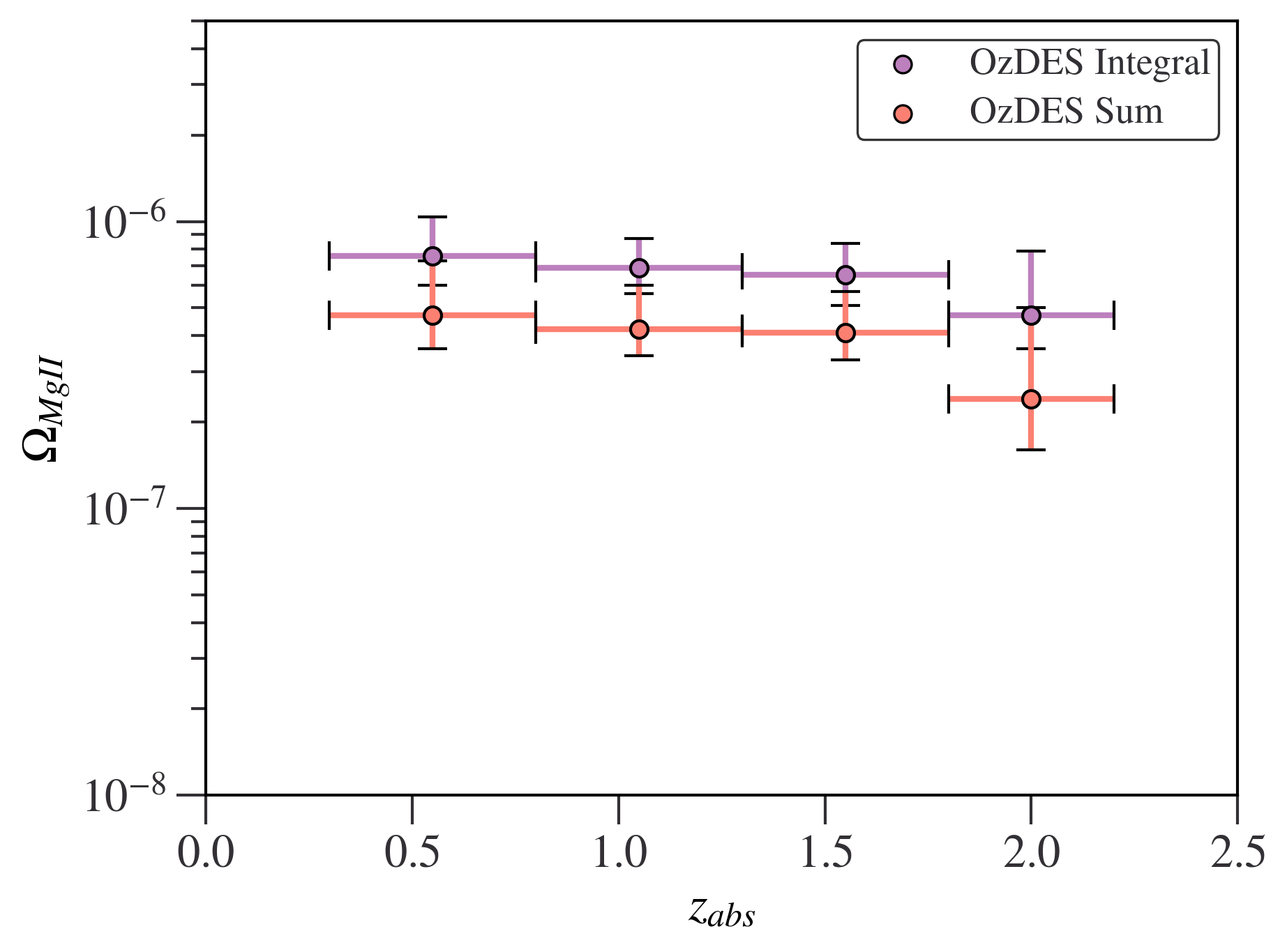}
\captionsetup{justification=justified, singlelinecheck=off} 
\caption{The mass density, $\Omega\subMgII$, for the the OzDES science sample ($W_r \geq 0.3$~{\AA}).  Values have been computed using both the summation formalism (Eq.~\ref{eqn:Omega Sum}, red points) and as an integration method (Eq.~\ref{eqn:Omega Integral}, purple points). }  
\label{fig:omegamg2}
\end{figure}

Over the redshift range $0.3 \leq z \leq 2.2$, there seems to be little-to-no evidence for evolution in the cosmic mass density of Mg$^{+}$ ion, though the mass density weakly trends toward higher values over the $\sim\! 7$~Gyr cosmic time corresponding to the studied redshift range. The integral values are marginally higher than the summation values because the integral method includes additional area under the column density distribution function at large column densities (we integrate to $N\subL=\infty$).  This results in a slightly larger mean column density,  $\langle N \rangle$, for the sample, as compared to the summation method, which terminates at the maximum column density of a realization.

\newpage
\section{Discussion}
\label{sec:discussion}

The analysis of intervening absorption lines is but one of many approaches to studying cosmological evolution.  Measuring evolution in the global star formation density provides insights into the average rate at which baryonic mass converts into stars across cosmic time \citep[e.g.,][]{reddy2009, oesch2013, oesch2014, mad-dick2014, bouwens2015}. Similarly, measuring evolution in the global energy density of ultraviolet radiation provides insights into the ionization history of the universe \citep[e.g.][]{schiminovich2005, reddy2009, oesch2013, oesch2014, bouwens2015}. Complimentary to such  measurements is the global mass density of gas, which provides insights into the cosmic evolution of the gas mass partaking in the baryon cycle of galaxies \citep[e.g.,][]{shull2012, nicastro2018, driver2021}. In particular, comparing evolution between the mass densities of various ions provides insights into both the chemical evolution of various chemical species and their relative ionization conditions.

Whereas the mass densities of {\HI} absorbers \citep[e.g.,][]{songaila2010, crighton2015, rao2017, ho2021}, of {\CIV} absorbers \citep[e.g.,][]{songaila2001, songaila2005, schaye2003, danforth2008, becker2009, ryan-weber2009, cooksey2010, dodorico2010, dodorico2013, simcoe2011, boksenberg2015, diaz2016, bosman2017, Codoreanu2018, manuwal2021, Davies2023}, and of {\SiIV} absorbers \citep[e.g.,][]{shull2014, cooksey2011, songaila2001, scannapieco2006, boksenberg2015, Codoreanu2018, dodorico2022} have been well documented, there are only three previous works that present measurements of $\Omega\subMgII$.  The first is \citet[][hereafter M17]{mathes2017}, who employed AOD column densities of 1180 {\MgII} absorbers with $0.003 \leq W_r \leq 8.5$~{\AA} covering $0.14 \leq z \leq 2.64$ measured in 602 HIRES and UVES quasar spectra.  The second is 
\citet[][hereafter C17]{Codoreanu2017}, who used Voigt profile column densities of 52 {\MgII} absorbers with $0.12 \leq W_r \leq 3.67$~{\AA} covering $2.0 \leq z \leq 5.4$ measured in four X-Shooter quasar spectra.  The third is \citet[][hereafter S23]{sebastian23}, who also used Voigt profile column densities of X-Shooter quasar spectra. The \citetalias{sebastian23} measurements were based on 280 {\MgII} absorbers in the redshift range $2.0 \leq z \leq 6.4$ identified in 42 spectra from the E-XQR-30 survey of \citep{dodorico23} as analyzed by \cite{Davies2023-Absorbers}. 

As described in Section~\ref{sec:massdensity}, 
we have developed a Monte Carlo method to estimate the mass density of {\MgII} absorbers, $\Omega\subMgII$, from equivalent widths measured in low-resolution spectra as informed by high-resolution spectra.  We have applied this method to the OzDES sample of {\MgII} absorbers over the redshift range $0.3 \leq z \leq 2.2$.  It is of interest to compare the OzDES {\MgII} mass density to the works \citetalias{mathes2017} and to examine evolution through comparison with the $2 \leq z \leq 6.5$ measurements by \citetalias{Codoreanu2017} and \citetalias{sebastian23}.

\subsection{The Mass Density at $0.3 \leq z \leq 2.2$}

As seen in Figure~\ref{fig:omegamg2}, the OzDES $\Omega\subMgII$ measurements indicate little-to-no evolution for $z \leq 2$.  This lack of strong evolution following Cosmic Noon is astrophysically plausible. Though $dN/dX$ of the strongest absorbers ($W_r>1.0$~{\AA}) decreases from Cosmic Noon to $z \sim 0.3$, it is no more than a $\sim\! 30$\% decline during this cosmic period \citep{Seyffert_13}. We note that the corresponding $dN/dX$ measured with OzDES is highly consistent with this very mild evolution (see Figure~\ref{fig:dNdz}(f)). Moreover, in this redshift regime (1) the ultraviolet background is declining by a two orders of magnitude \citep{haardt12, khaire19, Faucher-Giguere2020}, so that absorbing gas structures are evolving toward lower ionization conditions, and (2) the metallicity of {\MgII} absorbing structures (sub-Lyman limit systems and sub-damped {\Lya} absorber) are increasing by up to an order of magnitude \citep[e.g.,][]{fumagalli16, lehner23}.  Taken together, these statistical astronomical phenomena could yield an overall outcome of a more-or-less constant mass density over the redshift range $0.3 \leq z \leq 2.2$. 

\begin{figure*}[ht]
\centering
\includegraphics[width=0.9\linewidth]{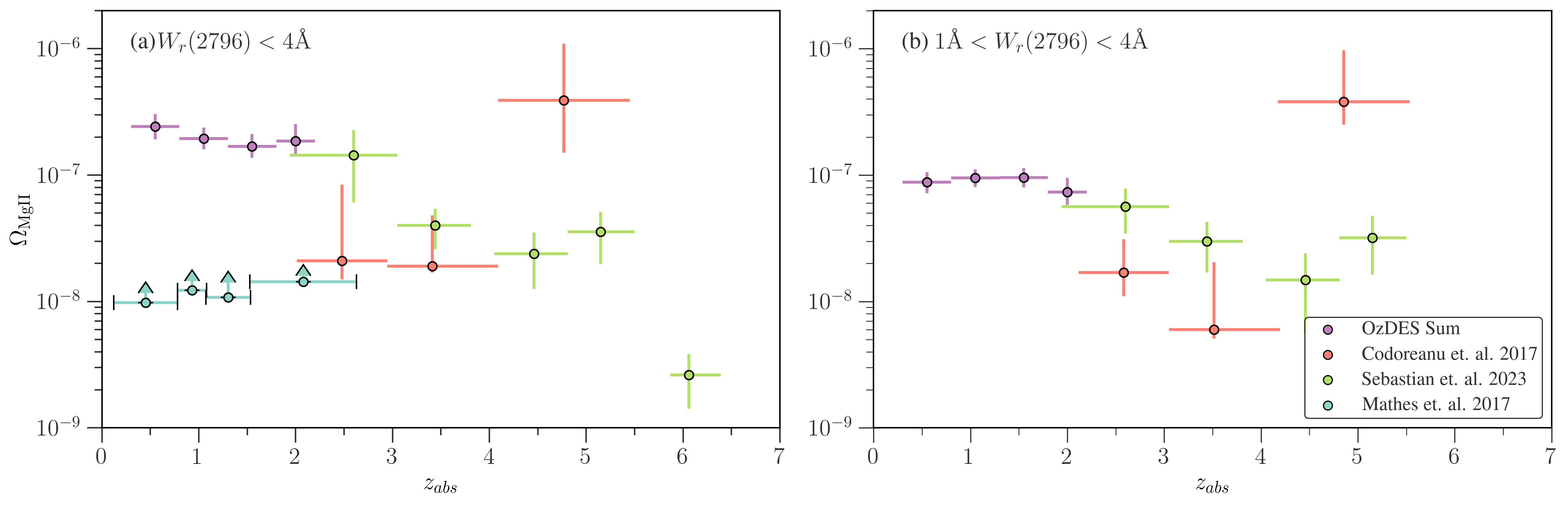}
\captionsetup{justification=justified, singlelinecheck=off} 
\caption{Comparison of redshift evolution of $\Omega\subMgII$ between the OzDES, \citetalias{mathes2017}, \citetalias{Codoreanu2017}, and \citetalias{sebastian23} surveys. The upper limit on the column density in the \citetalias{mathes2017} survey was $\log (N\subMgII/{\rm cm}^{-2}) \simeq 15$. (a) Absorbers with $W_r<4$~{\AA}.  The purple OzDES points are based on the summation method.  The green points ($z\geq 2$) are from \citetalias{Codoreanu2017}. The blue points are the upper limits from \citetalias{mathes2017} (note, however, that 7/1180 of their systems have $W_r > 4$~{\AA}). The orange points are from \citetalias{sebastian23}. (b) Absorbers limited to the range $1 \leq W_r < 4$~{\AA}. Data points are the same as in panel (a).}
\label{fig:Mega Omega}
\end{figure*}

In Figure~\ref{fig:Mega Omega}(a), we present a  comparison between the OzDES $\Omega\subMgII$ values and those reported by \citetalias{mathes2017} (orange points). The maximum equivalent width in the OzDES survey is $W_r = 3.5$~{\AA}, where as 7/1180 of the systems in the \citetalias{mathes2017} survey have $W_r > 4$~{\AA}. Both works cover the approximate redshift range $0.3 \leq z \leq 2.2$ in four similar redshift bins.   We present the OzDES values computed for both the summation (Eq.~\ref{eqn:Omega Sum}, purple points) and integral (Eq.~\ref{eqn:Omega Integral}, red points) methods. As  the values of \citetalias{mathes2017} are based on AOD column densities, and the AOD method provides only lower limits in the case of unresolved saturation \citep[e.g.,][]{savage_aod91, jenkins1996}, the \citetalias{mathes2017} $\Omega\subMgII$ values should be viewed as lower limits. 

As seen in Figure~\ref{fig:Mega Omega}(a), the OzDES $\Omega\subMgII$ values reside roughly a factor of ten (1~dex) above the $\Omega\subMgII$ lower limits of  \citetalias{mathes2017}.  Again, one clear reason for the substantially larger OzDES values is that fact the AOD method provides only lower limits for $\Omega\subMgII$ because of it provides lower limits on the {\MgII} column densities when the absorption profiles suffer unresolved saturation.  The largest  AOD column density lower limits in the \citetalias{mathes2017} sample are $\log (N\subMgII / {\rm cm}^{-2}) \simeq 15.0$.  Visual inspection of Figure~\ref{fig:log N vs log W} reveals that roughly 30\% of the absorbers in the \citetalias{Churchill_2020} sample have $\log (N\subMgII / {\rm cm}^{-2}) > 15.0$ for almost the complete range of {\MgII} rest-frame equivalent widths. Though the largest system-total column densities are $\log (N\subMgII / {\rm cm}^{-2}) \simeq 16.5$ from the Voigt profile fits of \citetalias{Churchill_2020}, the Monte Carlo method draws from a Schechter distribution function, which can yield even larger column densities for $W_r > 3$~{\AA} (though with a vanishing frequency). Thus, considering Eq.~\ref{eq:Omega_MgII}, which shows that the calculation of $\Omega\subMgII$ is proportional to the mean column density of the population, it is clear that the Monte Carlo method we applied to the OzDES {\MgII} systems would yield values that are  substantially larger than the lower limits of \citetalias{mathes2017}.

A similar outcome has been found for $\Omega\subCIV$, as demonstrated by \citet{cooksey2013}.  They used the SDSS DR7 quasar catalog \citep{schneider2010} and found $\sim\! 15,000$ {\CIV} absorbers in $\sim\! 26,000$ quasar sight lines.  They computed $\Omega\subCIV$ using AOD column densities, even though they were fully aware that the low-resolution SDSS spectra would yield a lower limit on $\Omega\subCIV$ in each redshift bin. In comparing their value to several high-resolution works that used Voigt profile column densities, they found that their AOD-based estimates of $\Omega\subCIV$ were systematically below those of the high-resolution works.

\subsection{The Mass Density at $0.3 \leq z \leq 6.4$}

Together, the OzDES, \citetalias{Codoreanu2017}, and \citetalias{sebastian23} surveys span Cosmic Noon and the epoch of {\HeII} reionization and probe the very end of the epoch of {\HI} reionization.

\citetalias{Codoreanu2017} conducted a survey of 
four quasars and found 52 intervening {\MgII} absorbers in the range $2 \leq z \leq 5.4$. The column densities of the {\MgII} absorbers were estimated using Voigt profile fitting. They computed $\Omega\subMgII$ in three redshift bins for two sample using the summation method (see Eq.~\ref{eqn:Omega Sum}). The first sample was for systems with $W_r \leq 4$~{\AA}, and the second was for systems with $1.0 \leq W_r \leq 4$~{\AA}.  The maximum system-total column density (i.e., sum of components) in their sample was $\log (N/{\rm cm}^{-2}) = 16.4$.  To ensure a self-consistent comparison between OzDES and \citetalias{Codoreanu2017}, we recomputed the OzDES $\Omega\subMgII$ using the equivalent width and column density ranges of \citetalias{Codoreanu2017}. \citetalias{sebastian23} adopted survey results of \citet{Davies2023-Absorbers}, which yielded 280 intervening {\MgII} absorbers in the range $1.9 \leq z \leq 6.4$ from 42 quasar spectra. The column densities of the {\MgII} absorbers were estimated using Voigt profile fitting. Using these data, \citetalias{sebastian23} computed $\Omega\subMgII$ in five redshift bins using the summation method (see Eq.~\ref{eqn:Omega Sum}) including all absorbers in their sample. 

In Figure~\ref{fig:Mega Omega}(a), we present $\Omega\subMgII$ for {\MgII} absorbers with $W_r<4$~{\AA} and, in Figure~\ref{fig:Mega Omega}(b), we present {\MgII} absorbers with $1 \leq W_r<4$~{\AA}.  We present the OzDES values computed for both the summation (Eq.~\ref{eqn:Omega Sum}) and integral (Eq.~\ref{eqn:Omega Integral}) methods. Both the \citetalias{Codoreanu2017} and \citetalias{sebastian23} were computed using the summation method. For the $W_r<4$~{\AA} sample, we also include the upper limits computed by \citetalias{mathes2017}, but we note that 7/1180 (0.6\%) of their systems have $W_r > 4$~{\AA}.

As seen in Figure~\ref{fig:Mega Omega}(a,b), the \citetalias{Codoreanu2017} values indicate a $\sim\! 1$ dex decline from $z\sim 5$ to $z\sim 3.5$ and are consistent with no evolution from between $z\sim 4$ to $z\sim 2$.  This would suggest a 10-fold decrease in the mass density of the the Mg$^+$ ion as traced by {\MgII} absorbers over a timescale of $\simeq\! 0.6$~Gyr (the approximate main sequence lifetime of a B8 or B9 star!)\footnote{The \citetalias{Codoreanu2017} sample is based on only four quasar spectra. The reason we believe that there is the factor of 10 difference is due to cosmic variance over this small sample.  The maximum column densities in the $z<4.5$ bins is $\log (N/{\rm cm}^{-2}) = 15.3$, whereas the maximum in the $z>4.5$ bin is 16.4. This neatly explains the factor of $\sim\! 10$ difference.}.  Taken at face value, the OzDES results would then imply a rapid increase of $\sim\! 1$~dex around $z\sim 2$. The OzDES results would constitute a second dramatic evolution over a period of $\sim\! 1$~Gyr that would be followed by no evolution from $z \sim 2$ down to $z\sim\! 0.3$ (roughly 7~Gyr).  

A different evolutionary picture emerges when we compared the OzDES mass densities to those measured by \citetalias{sebastian23}. As seen in Figure~\ref{fig:Mega Omega}(a,b), together these surveys would suggest a relatively smooth factor of 10 increase from $z\sim 5$ to Cosmic Noon followed by a plateau from $z\sim 2$ to $z \sim 0.4$ (we note that the highest redshift point of \citetalias{sebastian23} is the least secure and that no $W_r>1$~{\AA} absorbers were detected in this redshift bin). This smoother evolution would be a more astrophysically plausible evolution in that, as pointed out by \citetalias{sebastian23}, it is consistent with the factor-of-several increase in $dN/dX$ of the strongest {\MgII} absorbers from $z\sim 6$ to Cosmic Noon \citep{matejek-simcoe2013, chen-simcoe2017, sebastian23}.

\subsection{How Might the Mass Density Evolve?}

There are multiple ways in which $\Omega\subMgII$ could evolve: (1) metallicity evolution, (2) ionization evolution, (3) structure evolution, (4) all of the above. These astrophysical changes would manifest as evolution in the comoving redshift path density, which is the normalization to the column density distribution function. Alternatively, they could manifest as evolution in the shape of the column density distribution function.  Or, both the normalization and the shape could evolve. 

If the evolution is manifest in the normalization, $d{\cal N}/dX = (H_0/c)n_0\sigma_0$, then it would imply a change in either the comoving spatial number density, $n_0$, and/or the cross section of the absorbing structure, $\sigma_0$. Either could indicate chemical, ionization and/or structure evolution of the absorber. If the evolution is manifest in the shape of the column density distribution function, then it would imply changes in the conditions that create, sustain, and destroy smaller sized absorbers relative to larger sized absorbers (or vice versa). Evidence of shape evolution may even suggest that the types of astrophysical environments or overdensities populated by the {\MgII} absorbers evolve.

Evolution in the column density distribution function, whether due to evolution in its normalization, $d{\cal N}/dX$, and/or its functional shape, such as its power-law index $\alpha$, cannot directly inform us of the underlying astrophysics in the evolution of {\MgII} absorbing structures. At a minimum, deeper insights requires ionization modeling, which requires observations of multiple ions.  However, there are direct functional relationships between $\Omega\subMgII$, $d{\cal N}/dX$, and the column density distribution function such that their measurements should yield a self-consistent picture of {\MgII} absorbers.  In particular, such relationships have been exploited to measure the evolution in the equivlanet width distribution function from $z=7$ to $z=0$ \citep{churchill2024}. 

If the lack of evidence for evolution in the OzDES $\Omega\subMgII$ values for redshift following Cosmic Noon are taken at face value, we might infer that the various cosmic conditions governing the global mass density of {\MgII} absorbers have reached a steady state balance following the epoch of peak activity in the low-ionization gaseous component of the universe. 

\newpage
\subsection{Comparison to $\Omega\subHI$ and $\Omega_{\rm dust}$}

A comparison of $\Omega\subMgII(z)$ and $\Omega\subHI(z)$ can provide some insights into the redshift evolution of the ratio of the mean column densities of the population of {\MgII}-selected absorbers relative to that of the population of {\HI}-selected damped {\Lya} absorbers (DLAs), which are defined to have $\log (N\subHI/{\rm cm}^{-2}) \geq 20.3$ \citep{wolfe86}. From Eq.~\ref{eq:Omega_MgII} , we write
\begin{equation}
\frac{\langle N\subMgII \rangle}
     {\langle N\subHI \rangle} 
     = 
\frac{m_{\subH}}{m_{\subMg}}
\frac{\left( d{\cal N}/dX \right)\! \subHI}
     {\left( d{\cal N}/dX \right)\! \subMgII}
\frac{\Omega\subHI}{\Omega\subMgII} \, ,
\label{eq:Nbarrats}
\end{equation}
which can be evaluated as a function of redshift.  

In what follows, our estimates apply to $W_r \geq 1$~{\AA} {\MgII} absorbers. Adopting the collective measurements of \citet{Seyffert_13}, \citet{zhu_2013_the}, \citet{chen-simcoe2017}, and \citet{sebastian23} for $(d{\cal N}/dX)\subMgII$ over the redshift range $0.3 \!\leq\! z \!\leq\! 6.5$, and adopting the measurements of \citet{rao_2006_damped}, \citet{zafar13_PaperI, zafar13_PaperII} for $(d{\cal N}/dX)\subHI$ over the redshift range $0 \!\leq\! z \!\leq\! 5$, we can estimate the ratio of the comoving path densities in Eq.~\ref{eq:Nbarrats} over a wide range of redshift.  Similarly, we combine the $\Omega\subMgII$ measurements for $0.3 \!\leq\! z \!\leq\! 2$ from this work with those of \citet{sebastian23} for $2 \!\leq\! z \!\leq\! 6.5$ and adopt the fitted curve to $\Omega\subHI$ obtained from the meta-analysis of DLAs and 21-cm absorbers by \citet{peroux2020} to estimate the ratio $\Omega\subHI / \Omega\subMgII$ as a function of redshift.   

For $z\simeq 1$, we find that the ratio of the comoving path densities is $\sim 0.4$ and that the ratio of the mass densities is $\sim 10^{-4}$, and obtain
${\langle N\subMgII \rangle}/{\langle N\subHI \rangle} \simeq 1.6\!\times\!10^{-6}$. For $z\simeq 5$, we find that the ratios are $\sim 4$ and $\sim 2\!\times\! 10^{-6}$, respectively, and obtain
${\langle N\subMgII \rangle}/{\langle N\subHI \rangle} \simeq 3.3\!\times\!10^{-7}$. \citet[][see their Fig.~8]{rao_2006_damped} show that 
$\log \langle N\subHI \rangle / {\rm cm}^{-2} \simeq 21$, which would imply that
$\log \langle N\subMgII \rangle / {\rm cm}^{-2} \simeq 15.2$ at $z=1$ and 
$\log \langle N\subMgII \rangle / {\rm cm}^{-2} \simeq 14.5$ at $z=5$, an increase of roughly a factor of $\sim\! 5$ over the cosmic time spanning this redshift range.  This change provides a crude insight into the degree to which the metallicity and ionization conditions of $W_r \geq 1$~{\AA} {\MgII} absorbers evolve with cosmic time.  This simple exercise would suggest that metallicity is increasing and ionization level may be decreasing such that the mean column density of the population of strong {\MgII} absorbers increases by a factor of several across Cosmic Noon. It would seem plausible that constraints on the average metallicity evolution of DLAs could be measured using this method if the comoving path density and mass density of the {\MgII} absorbers applied in the ratios were accurately measured for DLA-selected absorption systems.

Dust can induce reddening on background quasars, the effect of which can be estimated using $\Omega_{\rm dust}$ in the {\MgII} bearing clouds.  
\citet{Menard_dust} measured $\Omega_{\rm dust}$ for $0.5<z<2$ {\MgII} absorbers and concluded that the slow build up of dust with cosmic time was consistent with long-lived {\MgII} clouds that are continually integrating dust from star formation of their host galaxies. Comparing to our measured $\Omega\subMgII$, we find the redshift evolution of $\Omega_{\rm dust}$ evolves in parallel with $\Omega\subMgII$ with a ratio $\Omega_{\rm dust}/\Omega\subMgII \sim 2$--3. This constant ratio is consistent with the inferences of \citet{Menard_dust} about the nature of {\MgII} absorbers.

\section{Conclusion}
\label{sec:conclusion}

We conducted a survey for intervening absorption lines from the  {\MgIIdblt} fine-structure doublet across the redshift range $0.3<z<2.2$ in 951 background quasar spectra obtained from the Data Release 2 of the Australian Dark Energy Survey (OzDES) \citet{lidman_2020_ozdes}.  The spectral resolution ranges from $R\simeq 1400$ to $R\simeq1700$ and the signal-to-ratios of the spectra range from 20--80, with more than half above 40.  We detected 337 {\MgII} absorbers with $W_r\geq 0.3$~{\AA} at the 50\% completeness level. This study comprises the first blind survey of {\MgII} absorbers conducted solely in the southern hemisphere (below the celestial equator).

We measured the redshift path density, $d{\cal N}/dz$, the comoving path density, $d{\cal N}/dX$, and the equivalent width distribution, $n(W_r) = dN/dXdW_r$, per unit absorption path per unit equivalent width.  We developed a method for estimating the cosmic {\MgII} mass density from the equivalent widths of the low-resolution {\MgII} lines. The model is informed by a high-resolution survey of {\MgII} absorbers for which column densities were measured using Voigt profile decomposition. 

We compared the OzDES southern hemisphere measurements of $d{\cal N}/dz$, $d{\cal N}/dX$, and $n(W_r)$ to those published from comprehensive {\MgII} surveys in the northern hemisphere across similar redshifts \citep[e.g.,][]{steidel1992, nestor_2005_mgiiabsorption, lundgren2009,  Seyffert_13, zhu_2013_the, Raghunathan2016}. We also compared our estimates for $\Omega\subMgII$ to the results of \citet{mathes2017} and \citet{Codoreanu2017}, the latter allowing us to perform a longitudinal study of redshift evolution from $z=5.5$ to $z=0.3$. 

Our main findings are as follows:

\begin{enumerate}

\item For {\MgII} absorbers in the southern hemisphere, the redshift path density, $d{\cal N}/dz$, and the comoving path density, $d{\cal N}/dX$, are consistent with the measured values in northern hemisphere surveys \citep[e.g.,][]{nestor_2005_mgiiabsorption, zhu_2013_the, Seyffert_13}. For $W_r>0.6$~{\AA} and $W_r> 1.0$~{\AA}, $d{\cal N}/dX$ is flat, suggesting no evolution over the $\sim\! 7$~Gyr period following Cosmic Noon.

\item The equivalent width distribution of the {\MgII} absorbers in the southern hemisphere is well fit by an exponential function with $N_* = 0.76\pm0.04$ and $W_* = 0.76\pm0.01$~{\AA}. These fitted parameters are consistent with those obtained from well-known northern hemisphere surveys \citep[e.g.,][]{nestor_2005_mgiiabsorption, Seyffert_13}. 

\item We develop a Monte Carlo model to compute the statistical properties of the column densities for {\MgII} absorbers in low resolution spectra from the rest-frame equivalent widths. We demonstrated that the model is highly successful at reproducing the {\MgII} column density distribution function measured from a high-resolution sample. This method can be generalized to other ions observed in low-resolution spectra.  
 
\item Estimation of {\MgII} column densities from absorber equivalent widths enabled us to compute the mass density, $\Omega\subMgII$, over $0.3\leq z \leq 2.2$. We obtained $\Omega\subMgII \sim 5 \times 10^{-7}$, roughly a factor of $\sim\! 10$ higher than the upper limits of \citet{mathes2017} over the same redshift range. The OzDES estimate is also roughly a factor of $\sim\! 10$ higher than the measurements of \citet{Codoreanu2017} for $2 \leq z \leq 4$, which would indicate a dramatic discontinuity in the evolution immediately following Cosmic Noon.  However, the OzDES measurements are consistent with a smooth flattening of the {\MgII} mass density after Cosmic Noon following the steady increase measured by \citet{sebastian23} from $z\sim 6$ to $z\sim 2$.

\end{enumerate}

It is well known that the calculation of the mass density is dominated by the largest column density systems. In a typical high-resolution {\CIV} survey, the path length covered is more than an order of magnitude smaller than that of, for example, the {\CIV} SDSS survey by \citet{cooksey2013}. Because of the steep column density distribution function, the highest column density absorbers are rare, and are likely not captured in the low-resolution surveys. This would bias their value of $\Omega\subCIV$ downward. In the other hand, as discussed by \citet{cooksey2013}, large low-resolution surveys have enough path length to find large column density systems, but the problem is that the column densities cannot be accurately determined from the low resolution spectra. This is the problem we faced with the OzDES survey.  We attempted to resolve this problem by using the high-resolution surveys to inform us of the statistical relationship between column density and equivalent width.  That we were able to use our model to accurately reproduce the column density distribution function of {\MgII} absorbers from their equivalent widths provides confidence that our method is robust.

On the other hand, we could test the validity of our method by applying it to a {\CIV} absorbers in the Sloan Digital Sky Survey (SDSS). These absorbers have been observed extensively with high-resolution spectra. The large database of Voigt profile fits and published $\Omega\subCIV$ values \citep[e.g.,][]{songaila2001, songaila2005, schaye2003, danforth2008, becker2009, ryan-weber2009, cooksey2010, dodorico2010, dodorico2013, simcoe2011, boksenberg2015, diaz2016, bosman2017, Codoreanu2018, manuwal2021, Davies2023} would serve as an excellent test of the Monte Carlo method.

This approach would allow us to validate our column density estimation method. It would build confidence that the model could be applied to any ion for which there is a large database of low resolution measurements and suitable (but smaller) high-resolution database to inform the model.  As such, the method holds promise for significantly increasing our ability to build a  census of the mass density of the multiple phases and astrophysical environments of the gaseous universe.

\section*{Acknowledgments}
The authors thank the anonymous referee for comments that helped improve the clarity of manuscript.  Much gratitude to Alma Sebastian for sharing their data prior to publication and especially for obliging our request to compute the data presented in Figure~\ref{fig:Mega Omega}(b) from their survey. A.A. acknowledges support from a Webber Fellowship administered through the Department of Astronomy at New Mexico State University.
G.G.K acknowledge the support of the Australian Research Council Centre of Excellence for All Sky Astrophysics in 3 Dimensions (ASTRO 3D), through project number CE170100013.

\newpage
\appendix

\section{The Column Densities Model}
\label{sec:appA}

The total column density of an absorber tends to increase with equivalent width in proportion to the number of Voigt profile components comprising the system \citep[e.g.,][]{petitjean1990, Churchill_2003, Churchill_2020}.  There is an element of scatter, however, as the equivalent with depends on the kinematics of the components, and the system-total column density depends on the column densities of the individuals components.  Thus, unfortunately, there is no one-to-one relationship; absorbers of a given $W_r$ can have a range of system-total column densities. We therefore employ a statistical approach and estimate the system-total $N$ for a given $W_r$ from the column density distribution function.  
As shown in Figure~\ref{fig:log N vs log W}(b), for the 198 high-resolution {\MgII} systems with $W_r \geq 0.3$~{\AA} that were Voigt profile fitted by \citetalias{Churchill_2020}, the system-total column density distribution function is well described by a Schechter function,
\begin{equation}
    f(N)dN = \Phi_0 \left( \frac{N}{N_*} \right)^{\alpha} \!\!
    \exp \left\{ -\frac{N}{N*} \right\}
    \frac{dN}{N_*}
\label{eq:Schchfit-All-appendix}
\end{equation}
for which we obtained  $\alpha=-1.16\pm0.03$  and $\log (N_*/{\rm cm}^{-2}) = 16.97\pm0.09 $ for the best-fitted parameters.  This serves as a global constraint for our model. That is, any ensemble of Monte Carlo realizations of the OzDES science sample must, on average, yield a measured column density distribution function consistent with the parameters we fitted to Eq.~\ref{eq:Schchfit-All-appendix} for the \citetalias{Churchill_2020} sample. 

\begin{figure}[ht] 
\centering
\includegraphics[width=0.95\columnwidth]{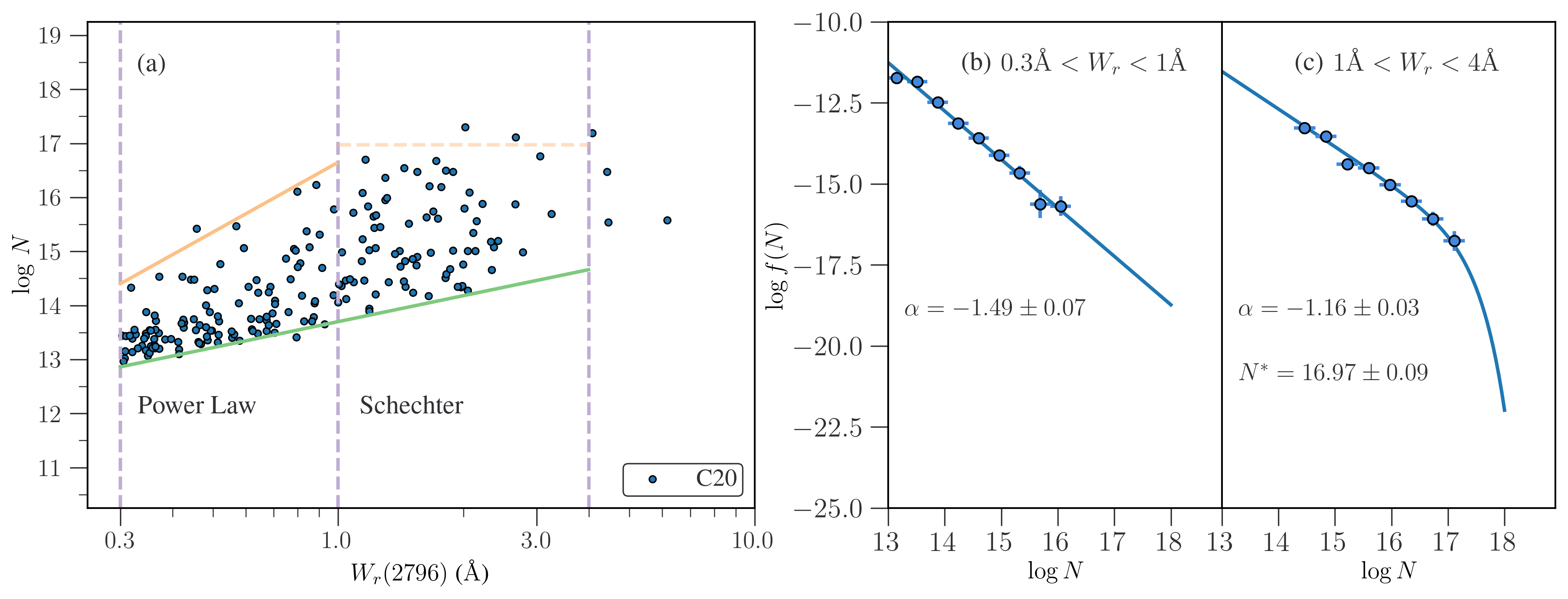}
\captionsetup{justification=justified, singlelinecheck=off} 
\caption{(a) The $N$--$W_r$ distribution of the \citetalias{Churchill_2020} data showing our estimates of the minimum (green line) and maximum (orange line) column density in different $W_r$ ranges. For $W_r \in (0.3,1.0]$~{\AA}, the column density distribution is modeled as a power law. For $W_r \in (1.0,4.0)$~{\AA}, we model the distribution as a Schechter function with no maximum column density. The horizontal dashed orange line corresponds to the characteristic value, $N_*$.  (b) The column density distribution for  $W_r \in (0.3,1.0]$~{\AA}. The fit over the range $\log (N/{\rm cm}^{-2}) \in (13.1,15.3)$ yields power-law index $\alpha=-1.49\pm0.07$. (c) The column density distribution for $W_r \in (0.6,1.0]$~{\AA}.  The adopted parameters over the range $\log (N/{\rm cm}^{-2}) \in (13.7,16.2)$ are $\alpha=-1.16\pm0.03$ and $\log (N_*/{\rm cm}^{-2}) = 16.97\pm 0.09$. }
\label{fig:NvsWcuts}
\end{figure}

However, we must also account for how the column density distribution function depends on $W_r$.  As can be seen in Figure~\ref{fig:NvsWcuts}(a), all systems with $W_r <1.0$~{\AA} in the \citetalias{Churchill_2020} sample have column densities smaller than the characteristic column density, i.e., $N<N_*$, and thus firmly reside on the power-law portion of the distribution function.  We fitted the power-law function $f(N) = \Phi_0 N^{\alpha}$ to the \citetalias{Churchill_2020} systems in the range $W_r \in (0.3,1.0]$ and obtained $\alpha =-1.49\pm 0.07$. The observed distribution and the fitted function are illustrated in Figure~\ref{fig:NvsWcuts}(b). For larger equivalent width systems in the range $W_r \in (1.0,4.0]$, column densities both below and above the characteristic column density, $N_*$, are represented. Thus, the column densities in this equivalent width range populate both the power-law and the exponential portion of the column density distribution function and are likely consistent with a populations drawn from a Schechter function. As shown in Figure~\ref{fig:NvsWcuts}(c), we fitted the \citetalias{Churchill_2020} systems in the range $W_r \in (1.0,4.0)$ and adopted $\alpha =-1.16\pm 0.03$ and $\log (N_*/{\rm cm}^{-2})=16.97\pm0.09$.

Furthermore, as can also be seen in Figure~\ref{fig:NvsWcuts}(a), the minimum, maximum, and range of column densities increases with increasing $W_r$. For example, systems with $W_r \simeq 0.3$~{\AA} have column densities in the range $\log (N/{\rm cm}^{-2}) \in (12.5,14.5)$, whereas systems with $W_r \simeq 1.0$~{\AA} have column densities in the range $\log (N/{\rm cm}^{-2}) \in (13.0,16.0)$, and those with $W_r \geq 1.0$~{\AA} exhibit $\log (N/{\rm cm}^{-2}) > 13.5$ with a rare few as high as $\log (N/{\rm cm}^{-2}) > 17$.  We parameterized the lower and upper boundaries for $N$ as a function of $W_r$ using simple linear functions of the form, $\log N(W_r) = N_0 + N_1 W_r$. We denote $N\subL(W_r)$ as the lower limit of the column density at $W_r$ and $N\subU(W_r)$ as the upper limit. Using the data in Figure~\ref{fig:NvsWcuts}(a), we found $N_0= 12.64$ and $N_1= 0.54$ for $N\subL(W_r)$ over the full range $W_r \in (0.3,4.0)$. For the upper limit, we limit the range for our parameterization of $N\subU(W_r)$ to $W_r \in (0.3,1.0]$, the region where the power-law distribution applies. We found $N_0 = 13.44$ and $N_1 = 3.21$. For $W_r \geq 1.0$~{\AA}, we allow $N\subL(W_r) = \infty$, as the exponential of the Schechter function provides the rapid decrease in the frequency of systems with $N>N_*$. The green curve in Figure~\ref{fig:NvsWcuts}(a) represents $N\subL(W_r)$, the orange curve represents $N\subU(W_r)$, and the horizontal dashed curve represents $N_*$ for the fit illustrated in panel~\ref{fig:NvsWcuts}(c).

As we describe below, we use Monte Carlo modeling to obtain realizations of the column densities for the OzDES science sample.  For $W_r \in (0.3,1.0]$~{\AA}, we draw column densities from the power-law distribution shown in Figure~\ref{fig:NvsWcuts}(b) in the range $N(W_r) \in (N\subL(W_r),N\subU(W_r))$ and, for $W_r \in (1.0,4.0]$~{\AA},  we draw from the Schechter distribution shown in Figure~\ref{fig:NvsWcuts}(c) in the range $N(W_r) \in (N\subL(W_r),\infty)$. 

\subsection{Power-law Regime}

For the range $W_r \in (0.3,1.0]$, we employ a power-law function of the form
\begin{equation} 
    f(t;\alpha) dt = t^{\alpha} dt \, ,
\label{eqn:plaw Freq Dist}
\end{equation}
where $\alpha$ is the power-law slope and $t$ represents the column density. As we show in Figure~\ref{fig:NvsWcuts}(b), we obtained $\alpha = -1.49\pm 0.07$ from a power-law fit to the \citetalias{Churchill_2020} data for this $W_r$ range. This steeper power-law in this equivalent width regime is an important component of our model, which otherwise overproduced large column densities for smaller $W_r$ systems. 

For a given system of $W_r$ in the OzDES sample, we assume that a random deviate, $x$, corresponds to the fractional area under the column density cumulative distribution function according to
\begin{equation}
    x = \displaystyle \frac{\displaystyle \int_{N\subL}^{N} \!\! t^{\alpha} \,dt}{\displaystyle \int_{N\subL}^{N\subU} \!\!\!\! t^{\alpha} \,dt} =
    \frac{N^{\alpha\!+\!1}-N\subL^{\alpha\!+\!1}}{N\subU^{\alpha\!+\!1} - N\subL^{\alpha\!+\!1}}
    \, ,
\label{eq:x2giveN}
\end{equation}
where $N \in (N\subL, N\subU)$, where $N\subL = N\subL(W_r)$ and  $N\subU = N\subU(W_r)$ are the lower and upper ranges to the column density distribution at $W_r$.  To obtain the column density corresponding to a given $x$, we rearrange Eq.~\ref{eq:x2giveN},  giving
\begin{equation} 
    N = \bigg[x\left(N\subU^{\alpha\!+\!1} - N\subL^{\alpha\!+\!1}\right) + N\subL^{\alpha\!+\!1}\bigg]^{1/\alpha\!+\!1} \, .
\label{eqn:plaw N estimate}
\end{equation}

\subsection{Schechter Regime}

For the range $W_r \in (1.0,4.0)$, we employ a Schechter function \citep{schechter76},
\begin{equation}
    f(t;\alpha) dt = t^{\alpha} \exp \left\{ -t \right\} dt \, ,
\end{equation}
where $t = N/N_*$. As we showed in Figure~\ref{fig:NvsWcuts}(c), we obtained $\alpha = -1.16\pm 0.03$ and $N_*=16.97\pm0.09$ from a Schechter function fit to the full \citetalias{Churchill_2020} sample with $W_r \geq 0.3$~{\AA}. 
For a given a system with a given $W_r$ in the OzDES sample, we assume that a random deviate, $x$, corresponds to the fractional area under the column density cumulative distribution function according to
\begin{equation}
  x = \frac
  {\displaystyle \int_{t\subL}^{t} t^{\alpha} \exp \left\{ -t \right\} dt}
  {\displaystyle \int_{t\subL}^{t\subU} t^{\alpha} \exp \left\{ -t \right\} dt}
 =   \frac {\displaystyle \Gamma(\alpha\!-\!1,t\subL) - \Gamma(\alpha\!-\!1,t)}
  {\displaystyle \Gamma(\alpha\!-\!1,t\subL) - \Gamma\alpha\!-\!1,t\subU)}
  \, ,
\end{equation}
where $\Gamma(\alpha\!-\!1,t)$ is the incomplete $\Gamma$ function.  Using Brent's method \citep{brent73}, we root solve 
\begin{equation}
\Gamma(\alpha\!-\!1,t) + 
  (x-1)\Gamma(\alpha\!-\!1,t\subL) - x \Gamma(\alpha\!-\!1,t\subU) 
  = 0 
  \, ,
\end{equation}
for $t$, from which we obtain $N=N_* t$.

\section{Assessing the Model}
\label{sec:appB}

The distribution and median $\Omega\subMgII$ obtained from our Monte Carlo modeling strongly depend on the ability of our model to generate realistic distributions of column densities. To evaluate the degree to which the Monte Carlo model captures the observed distribution of $N$--$W_r$ pairs of the \citetalias{Churchill_2020} sample, we performed $10^5$ trials using the equivalent widths of the \citetalias{Churchill_2020} systems. We compared each Monte Carlo realization to the observed distribution presented in Figure~\ref{fig:log N vs log W} using a 2D Kolmogorov–Smirnov (KS) test. We then examined the distribution of the KS-statistic probabilities ($p$-values) that the null hypothesis is satisfied, where the null hypothesis is that the two distributions are drawn from the same parent population.

\begin{figure}[ht] 
\centering
\includegraphics[width=\columnwidth]{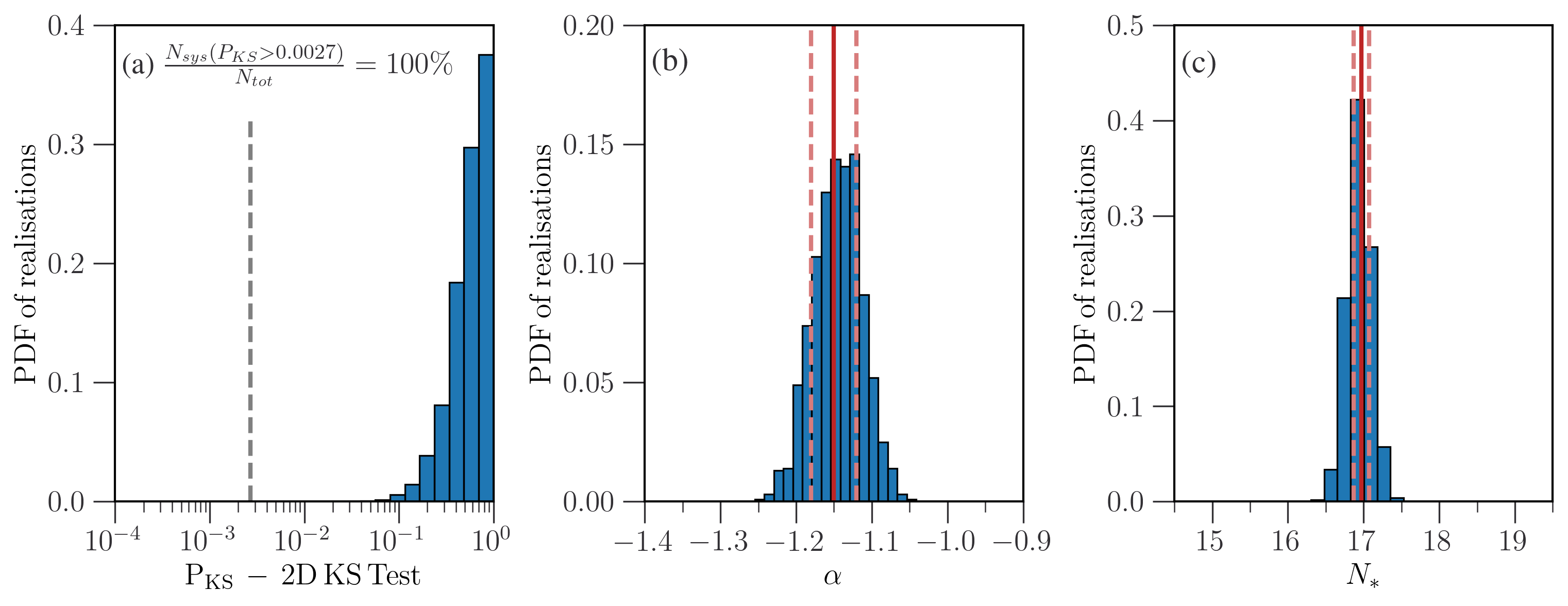}
\captionsetup{justification=raggedright}
\captionsetup{justification=justified, singlelinecheck=off} 
\caption{a) The distribution of the 2-dimensional KS $p$-values obtained by comparing $10^5$ Monte Carlo realizations of $N$--$W_r$ pairs to the observed distribution for the \citetalias{Churchill_2020} sample (see Figure~\ref{fig:log N vs log W}(a)). The black vertical line represents $p = 0.0027$, corresponding to a $99.73\%$ ($3\sigma$) confidence level. A $p$-value less than 0.0027 would suggest that the Monte Carlo realization is not drawn from the same parent population as the \citetalias{Churchill_2020} sample. No realization yielded $p<0.0027$.  The bulk of the KS tests yield $p>0.1$, indicating that the Monte Carlo realizations are highly consistent with the \citetalias{Churchill_2020} sample. (b,c) Distribution of fitted parameters $\alpha$ and $N_*$ of the column density distribution functions for $10^5$ realizations of the OzDES science sample. (b) The distribution of Monte Carlo fitted power-law indices (blue histogram) has mean and standard deviation $\alpha = -1.14\pm 0.05$, which is highly consistent with the measured value $\alpha=-1.16\pm0.03$ (red vertical lines) from the \citetalias{Churchill_2020} sample. (c) The distribution of Monte Carlo fitted characteristic column densities (blue histogram) has mean and standard deviation $\log (N_*/{\rm cm}^{-2}) = 16.93\pm0.23$, which is highly consistent with the measured value $\log (N_*/{\rm cm}^{-2}) = 16.97\pm 0.09$ (red vertical lines) from the \citetalias{Churchill_2020} sample.}  
\label{fig:KSTest-MCrealz}
\end{figure}

The distribution of KS $p$-values from $10^5$ Monte Carlo realizations is presented in Figure~\ref{fig:KSTest-MCrealz}(a). The vertical dashed line shows $p = {\rm erf}(3/\sqrt{2}) = 0.0027$.  For an individual realization, if $p \leq 0.0027$, we can reject the null hypothesis that the two compared samples are drawn from the same underlying parent population at a confidence level 99.73\% ($3\sigma$) or higher. We might expect that a non-zero fraction of the $10^5$ realizations will have $p \leq 0.0027$. If that fraction is 0.27\% (270 realization) or larger, we can reject the null hypothesis for the full suite of Monte Carlo realizations at the 99.73\% confidence level or higher. As shown in Figure~\ref{fig:KSTest-MCrealz}(a), none of the Monte Carlo realizations yielded $p \leq 0.0027$, which suggest we cannot rule out the null hypothesis, i.e., we can conclude that the $N$--$W_r$ distributions generated by the Monte Carlo model are consistent with the observed distribution of $N$--$W_r$ pairs of the \citetalias{Churchill_2020} sample.  In fact, the mode of the $p$-value distribution is $p\simeq 0.85$ and the vast majority of the KS tests yielded $p>0.1$. The distribution of $p$-values strongly indicate that our model consistently generates realizations that reflect the observed distribution of $N$--$W_r$ pairs for the \citetalias{Churchill_2020} sample. 

A second test is the to see how well the Monte Carlo realizations recover the column density distribution of the \citetalias{Churchill_2020} sample, which we measured to have a best-fit power-law slope  $\alpha = -1.16\pm 0.03$ and characteristic column density $\log (N/{\rm cm}^{-2}) = 16.97\pm 0.09$. We generated $10^5$ Monte Carlo realizations of the \citetalias{Churchill_2020} sample. For each realization of the sample, we obtained the best-fit power-law slope and characteristic column density assuming a Schechter function for the column density distribution. In Figure~\ref{fig:KSTest-MCrealz}(b), we plot the distribution of best-fit $\alpha$ values (blue histogram), which has mean and standard deviation $\alpha = -1.14\pm 0.05$. The red vertical lines represent the best-fit value and uncertainty from the \citetalias{Churchill_2020} sample.  In Figure~\ref{fig:KSTest-MCrealz}(c), we plot the distribution of best-fit $N_*$ values (blue histogram), which has mean and standard deviation $\log (N_*/{\rm cm}^{-2}) = 16.93\pm0.23$.  The red vertical lines represent is the best-fit value and uncertainty from the \citetalias{Churchill_2020} sample.  The high degree of consistency between the mean $\alpha$ and $N_*$ from the Monte Carlo generated column densities and that of the \citetalias{Churchill_2020} sample strongly suggest that our Monte Carlo realizations  accurately reproduce column density distribution functions consistent with observations.

\section{The Mass Density Calculation}
\label{sec:appC}

For a given Monte Carlo realization of the OzDES science sample, the mass density is computed from
\begin{equation} 
\Omega\subMgII (z)  = \frac{H_0}{c} \frac{m_{\hbox{\tiny Mg}}}{\rho_c} \langle N \rangle \frac{d{\cal N}}{dX} \, ,
\label{eq:Omega_MgII-appendix}
\end{equation}
where $\langle N \rangle$ is the measured mean column density of the sample of {\MgII} absorbers, and $d{\cal N}/dX$ is their comoving path density.  The product $\langle N \rangle d{\cal N}/dX$ is equivalent to the total column density per unit of comoving path length. It can be computed using the summation $\langle N \rangle d{\cal N}/dX = (\sum_i \! N_i)/\Delta X$, where the sum is taken over all absorbers in a given sample and $\Delta X$ is the comoving path length for the sample. Alternatively, since $d{\cal N}/dX$ is measured for the sample, we can compute (also see Eq.~\ref{eqn:Omega Integral})
\begin{equation} 
  \langle N \rangle 
 \frac{d{\cal N}}{dX}   =  
  \displaystyle\int^{N\subU}_{N\subL} \!\! f(N)NdN \, ,
\label{eqn:Omega Integral Appendix}
\end{equation}
from the first moment of the column density distribution function, $f(N)$, where $N\subL$ and $N\subU$ are the minimum and maximum column density for the sample.

For each Monte Carlo realization of the OzDES science sample, we computed $\alpha$ and $N_*$ from a fit to the distribution of column densities generated using the methods describe in Appendix~\ref{sec:appA}. To determine the mean column density of the realization, we perform the integral 
\begin{equation}
  \langle t \rangle = \frac
  {\displaystyle \int_{t\subL}^{\infty} t^{\alpha\!+\!1} \exp \left\{ -t \right\} dt}
  {\displaystyle \int_{t\subL}^{\infty} t^{\alpha} \exp \left\{ -t \right\} dt}
 =   \frac {\displaystyle \Gamma(\alpha,t\subL)}
  {\displaystyle \Gamma(\alpha\!-\!1,t\subL)}
  \, ,
\label{eq:tbar}
\end{equation}
where $t\subL = N\subL/N_*$ 
is the minimum and maximum column densities for the realization. We obtain $\langle N \rangle$ for the realization from $\langle t \rangle = \langle N \rangle/ N_*$.

\begin{figure}[ht] 
\centering
\includegraphics[width=1\columnwidth]{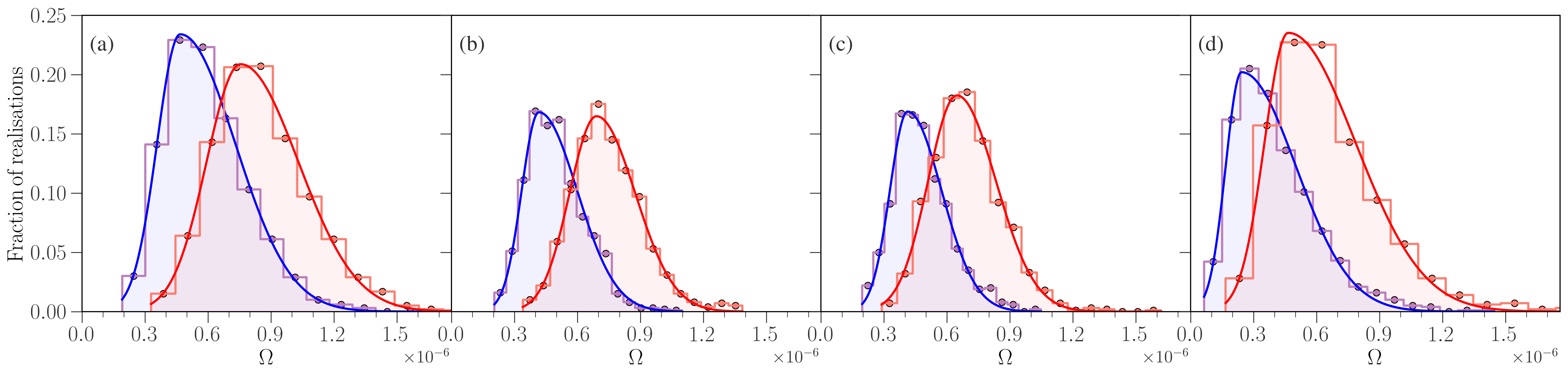}
\captionsetup{justification=raggedright}
\captionsetup{justification=justified, singlelinecheck=off} 
\caption{The binned distribution of $\Omega\subMgII$ for the OzDES science sample from the Monte Carlo modeling for four redshift ranges. Both the summation method (purple) and the integral method (red) are shown. (a) $z \in (0.33,0.80]$. (b) $z \in (0.80,1.30]$. (c) $z \in (1.30,1.80]$. (d) $z \in (1.80,2.20]$.   The quoted best values and their uncertainties are determined by fitting univariate asymmetric Gaussians (see Eq.~\ref{eq:uag}). The fits are shown by the curves through the binned distributions. The quoted values (see Eq.~\ref{eq:Omega-quote}) are presented in in Table~\ref{tab:omegamg2} and Figure~\ref{fig:omegamg2}.} 
\label{fig:Omegadist}
\end{figure}

We then compute $\Omega\subMgII$ for the realization using Eq.~\ref{eq:Omega_MgII} (also see Eq.~\ref{eq:Omega_MgII-appendix}) employing both the summation (Eq.~\ref{eqn:Omega Sum}) and integral (Eqs.~\ref{eqn:Omega Integral}, also see Eq.~\ref{eqn:Omega Integral Appendix}) methods.  In this way we have two estimates for $\Omega\subMgII$ for each realization. The quoted best value and its uncertainty was obtained by modeling the $\Omega\subMgII$ distributions using a univariate asymmetric Gaussian function \citep{kato02},
\begin{equation}
    f(\Omega;\Omega_0,\sigma,\beta) = 
    \frac{1}{\sqrt{2\pi}\,\sigma}
    \frac{2}{(1+\beta)} 
\begin{cases} 
\displaystyle
\exp \left[\frac{-(\Omega-\Omega_0)^2}{2\sigma^2} \right] & \, \Omega > \Omega_0 \\[12pt]
\displaystyle
\exp \left[\frac{-(\Omega-\Omega_0)^2}{2\beta^2 \sigma^2}\right] &  \, \Omega \leq \Omega_0\, , \\ 
\end{cases}
\label{eq:uag}
\end{equation}
where $\Omega_0$ is the most probable value, $\beta$ is the asymmetry parameter, $\sigma$ represents $34$\% of the area above $\Omega_0$, and $\beta\sigma$ represents $34$\% of the area below $\Omega_0$.  The mass densities are then quoted as 
\begin{equation}
\Omega\subMgII = \Omega_0\phantom{} ^{+\sigma}_{-\beta\sigma} \, .
\label{eq:Omega-quote}
\end{equation}

In Figure~\ref{fig:Omegadist}, we show the distributions of $\Omega\subMgII$ for $10^5$ Monte Carlo realizations of the OzDES science sample. Treating each realization as we would treat the real-world OzDES sample, we examine $\Omega\subMgII$ in four redshift bins for each realization. The purple distributions are for the summation method and the red distributions are for the integral method. The solid curves are the fitted functions given by Eq.~\ref{eq:uag}. The final quoted $\Omega\subMgII$ values are listed in Table~\ref{tab:omegamg2} and presented in Figure~\ref{fig:omegamg2}.

\bibliography{main}{}
\bibliographystyle{aasjournal}

\end{document}